\documentclass[preprint,prd,showpacs,superscriptaddress,amsmath,nofootinbib]{revtex4}

\usepackage{cancel}
\usepackage{graphicx}
\usepackage{color}
\usepackage{amssymb}

\newlength{\nseparation}
\newenvironment{nfigure}[1]
        {\begin{figure}[#1]\hrule\vspace{\nseparation}\par}
        {\vspace{\nseparation}\par \hrule \end{figure}}

\newcommand{\bea}{\begin{eqnarray}}
\newcommand{\eea}{\end{eqnarray}}
\newcommand{\nn}{\nonumber}

\newcommand{\eq}[1]{Eq.~(\ref{#1})}
\newcommand{\gev}{\,\textrm{GeV}}
\newcommand{\Msusy}{M_{\textrm{SUSY}}}
\newcommand{\CKM}{\textrm{CKM}}
\newcommand{\ps}{p\hspace{-0.44em}/\hspace{0.06em}}

\def\slashed#1{\displaystyle{\not}#1}

\begin{document}

\title{Complete resummation of chirally-enhanced loop-effects in
  the MSSM with non-minimal sources of flavor-violation\bigskip}

\author{Andreas Crivellin} \email{crivellin@itp.unibe.ch}
\affiliation{Albert Einstein Center for Fundamental Physics, Institute
  for Theoretical Physics,\\ University of Bern, CH-3012 Bern,
  Switzerland.\bigskip}

\author{Lars Hofer}\email{lars.hofer@physik.uni-wuerzburg.de}
\affiliation{Institut f\"ur Theoretische Physik und Astrophysik,
  \\ Universit\"at W\"urzburg, D-97074 W\"urzburg, Germany\bigskip}
             
\author{Janusz Rosiek\bigskip} \email{Janusz.Rosiek@fuw.edu.pl}
\affiliation{Institute of Theoretical Physics, Physics Department,
  University of Warsaw, Ho{\.z}a 69, 00-681 Warsaw, Poland \bigskip\bigskip}

\date{\today\bigskip\bigskip}

\begin{abstract}
In this article we present the complete resummation of the leading
chirally-enhanced corrections stemming from gluino-squark,
chargino-sfermion and neutralino-sfermion loops in the MSSM with
non-minimal sources of flavor-violation.  We compute the finite
renormalization of fermion masses and the CKM matrix induced by
chirality-flipping self-energies.  In the decoupling limit $\Msusy\gg
v$, which is an excellent approximation to the full theory, we give
analytic results for the effective gaugino(higgsino)-fermion-sfermion
and the Higgs-fermion-fermion vertices.  Using these vertices as
effective Feynman rules, all leading chirally-enhanced corrections can
consistently be included into perturbative calculations of Feynman
amplitudes.  We also give a generalized parametrization for the bare
CKM matrix which extends the classic Wolfenstein parametrization to
the case of complex parameters $\lambda$ and $A$.
\end{abstract}

\pacs{11.30.Pb,12.15.Ff,12.60.Jv,14.80.Da}

\maketitle

\section{Introduction}
\label{sec:intro}

In the Standard model (SM) left- and right-handed fermion fields $f_L$
and $f_R$ transform differently under the $SU(2)_L$ gauge symmetry.
Thus, the requirement of gauge invariance forbids explicit mass-terms.
Instead these fields acquire masses via the Higgs mechanism.  The
Higgs field $H$ (which is itself a $SU(2)_L$ doublet) couples
left-handed fermions to right-handed ones with coupling strength
$Y^{f_i}$ (i denotes the generation of the fermion), so that the
non-vanishing vacuum expectation value\footnote{We define
  $\left\langle H \right\rangle= v$ (without a factor $\sqrt{2}$), so
  that $v\approx 174\gev$.}  (vev) $v$ of $H$ then induces fermion
masses $m_{f_i}=Y^{f_i} v$.  Experimental measurements revealed
$m_{f_i}\ll v$ for all the fermions except for the top quark implying
$Y^{f_i}\ll 1\;(f_i\not= t)$.  Since the Yukawa couplings
$Y^{f_i}\;(f_i\not= t)$ are thus small compared to the gauge
couplings, their values are in principle sensitive to loop corrections
if such higher-order contributions manage to escape the
$Y^{f_i}$-suppression.  However, any loop correction to $Y^{f_i}$ has
to involve a chirality-flip and since in the SM the Yukawa couplings
are the only sources of chirality-violation the loop must be
proportional to $Y^{f_i}$ itself, so that the $Y^{f_i}$-suppression
cannot be avoided.\medskip

In the Minimal Supersymmetric Standard Model (MSSM) the situation is
different. Firstly, it contains two Higgs doublets $H_u$ and $H_d$
coupling to up- and down-type quark (lepton) superfields,
respectively.  The neutral components of these Higgs fields acquire
vevs $v_u$ and $v_d$ with $v_u^2+v_d^2=v^2$.  If there is a hierarchy
$v_d\ll v_u$, one faces enhanced corrections to Feynman amplitudes in
which the tree-level contribution is suppressed by the small vev $v_d$
while the loop correction involves $v_u$ instead.  In this case the
ratio of one-loop to tree-level contribution receives an enhancement
factor $\tan\beta\equiv v_u/v_d$ \cite{Hall:1993gn}.  Secondly, the
MSSM offers another source of chirality-flips, namely the soft
SUSY-breaking trilinear Higgs-sfermion couplings $A^f$ ($A$-terms)
with mass dimension one\footnote{Strictly speaking, the flip of
  fermion chirality is provided by a gaugino propagator in the
  corresponding loop diagram. However, the $A$-terms change the
  $SU(2)_L$\,-\,charge on the sfermion-line and in this sense they are
  also necessary in order to mediate the chirality-flip of the
  fermions.}.  Whereas one has $A^f\propto Y^f$ in a scenario of
Minimal Flavor Violation (MFV) \cite{D'Ambrosio:2002ex}, in the
general MSSM these couplings are independent free parameters.  Thus,
enhanced corrections to Feynman amplitudes in which the tree-level
contribution is suppressed by a small $Y^f$ while the loop correction
involves $A^f$ instead are possible.  In such a case the ratio of the
one-loop to the tree-level contribution receives an enhancement factor
$A^f_{ij}/(Y^f_{ij} \Msusy)$, where $\Msusy$ is a typical
SUSY-mass.\medskip

In both cases the respective enhancement factor ($\tan\beta$ or
$A^f_{ij}/(Y^f_{ij}\Msusy)$) can compensate for the loop suppression.
Therefore such a higher loop correction can be of the same size, or
even larger, as the leading order diagram\footnote{Since self-energy
  diagrams involving $A^f$-terms can be of the same order as the light
  fermion masses, they can even generate them entirely in a scenario
  with loop-induced soft Yukawa couplings \cite{Lahanas:1982et,
    Borzumati:1999sp, Ferrandis:2004ri, AJU}} and perturbative
calculations (using the usual counting in powers of $\alpha_s$,
$\alpha_{1,2}$) should thus be supplemented by an all-order
resummation of the enhanced corrections. In nearly all cases this can
be achieved by using effective Feynman rules which incorporate the
resummed corrections.  Such effective rules have already been
calculated in the literature for several special cases and vertices
$(S^0=H^0,A^0,h^0)$:
\begin{itemize}
  \item $S^0b\bar{b}$ and $H^+t\bar{b}$ vertices for $A^b=0$
    \cite{Hall:1993gn,Carena:1999py}.
  \item $S^0d_i\bar{d}_j$ and $H^+u_i\bar{d}_j$ vertices in the MSSM
    with MFV \cite{Hamzaoui:1998nu, Babu:1999hn,
      Isidori:2001fv,Buras:2001mb,Buras:2002vd,Buras:2002wq}.
  \item $S^0d_i\bar{d}_j$ and $H^+u_i\bar{d}_j$ vertices in the MSSM
    with MFV and additional sources of CP violation
    \cite{Dedes:2002er, Ellis:2007kb,
      Ellis:2009di}\footnote{Ref.~\cite{Ellis:2009di} also extends the
      analysis to general soft-SUSY-breaking terms by expanding them
      in terms of the Yukawa couplings.}.
  \item $S^0f_i\bar{f}_j$ and $H^+f_i\bar{f}^{\prime}_j$ vertices for
    quarks and leptons in the general MSSM in the limit $A^f=0$
    \cite{Chankowski:2000ng,Isidori:2002qe}.
  \item $S^0b\bar{b}$ vertex for $A^b\neq0$ including NNLO QCD
    corrections \cite{Guasch:2003cv}.
  \item $\tilde{g}\tilde{d}_i\bar{d}_j$,
    $\widetilde{\chi}^+\tilde{f}_i\bar{f}_j^{\prime}$,
    $\widetilde{\chi}^0\tilde{f}_i\bar{f}_j$ vertices for quarks and
    leptons in the general MSSM for $A^f=0$ \cite{Hisano:2008hn}. This
    method requires iterative resummation.
  \item Complete set of $S^0f_i\bar{f}_j$, $H^+f_i\bar{f}^{\prime}_j$,
    $\tilde{g}\tilde{d}_i\bar{d}_j$,
    $\widetilde{\chi}^+\tilde{u}_i\bar{d}_j$,
    $\widetilde{\chi}^-\tilde{d}_i\bar{u}_j$,
    $\widetilde{\chi}^0\tilde{d}_i\bar{d}_j$ vertices in the MSSM with
    MFV beyond the decoupling limit $\Msusy\gg v$ \cite{Hofer:2009xb}.
  \item Effective $\widetilde{\chi}^+\tilde{\ell}_i\bar{\nu}_j$ and
    $\widetilde{\chi}^0\tilde{\ell}_i\bar{\ell}_j$, vertices in the
    general MSSM (with $A^\ell=0$) beyond the decoupling limit
    $\Msusy\gg v$ \cite{Girrbach:2009uy}.
  \item $\tilde{g}\tilde{d}_i\bar{d}_j$,
    $\widetilde{\chi}^+\tilde{u}_i\bar{d}_j$,
    $\widetilde{\chi}^-\tilde{d}_i\bar{u}_j$ vertices in the general
    MSSM beyond the decoupling limit $\Msusy\gg v$ (corrections from
    gluino-squark loops only)
    \cite{Crivellin:2008mq,Crivellin:2009ar}.
  \item $S^0d_i\bar{d}_j$ and $H^+u_i\bar{d}_j$ vertices in the
    general MSSM including $A$-terms, $A^\prime$ terms beyond leading
    order in $v/\Msusy$ (corrections from gluino-squark loops only)
    \cite{Crivellin:2010er}.
\end{itemize} 

However, a complete list of the gaugino(higgsino)-fermion-sfermion
and Higgs-fermion-fermion vertices including the full set of chirally
enhanced corrections is still missing.  In this article we deliver the
missing pieces taking into account enhanced contributions from
gluino-squark, chargino-sfermion and neutralino-sfermion loops in the
general MSSM.  For the resummation we rely on the methods developed in
Refs.~\cite{Carena:1999py, Hofer:2009xb, Crivellin:2008mq,
  Crivellin:2009ar}, which can be applied for an arbitrary value of
the SUSY mass scale $\Msusy$, in particular beyond the decoupling
limit $\Msusy\gg v$.  In general, however, the resummation of
self-energy corrections requires iterative procedures and the enhanced
vertex corrections to the Higgs-fermion-fermion vertex cannot be
absorbed into an effective coupling.  These complications do not
occur if contributions which are subleading in $v/\Msusy$ are
neglected.  We present analytical resummation formulae in this limit,
which for realistic values of SUSY masses turn out to be an excellent
approximation to the full result: according to the new results
   of the CMS collaboration \cite{Collaboration:2011wc} and the
   Atlas experiment \cite{daCosta:2011qk}, squarks and
   gluinos must be rather heavy so that decoupling effects 
in squark-mixing can be neglected to a good approximation. Since $m_{\tau}<m_{b}$ and the off-diagonal $A^\ell$-terms  are severely constrained from experiments searching for flavor transitions in the 
charged lepton sector, the LR-elements in the slepton mass matrices are typically 
smaller than the ones in the squark mass matrices and decoupling effects in 
slepton-mixing can only be important if the sleptons are much lighter than the squarks.
Furthermore, chargino- and neutralino-mixing effects are suppressed by 
$M_W^2 / M_{{\rm SUSY}}^2$ and can be neglected if only $M_{{\rm 
SUSY}} > M_W$. These facts support our statement that the decoupling limit is almost always an excellent approximation.
\medskip

The paper is organized as follows.  In Sec.~\ref{sec:Self-energies} we
calculate the chirally-enhanced parts of the quark and lepton
self-energies in the MSSM.  Sec.~\ref{sec:Renormalization} is devoted
to the renormalization of Yukawa couplings, fermion wave-functions and
the CKM matrix in the presence of chirally-enhanced corrections.  Our
main result, the effective gaugino(higgsino)-fermion-sfermion and
Higgs-fermion-fermion vertices are presented in
Sec.~\ref{sec:effective-vertices} and Sec.~\ref{sec:higgs_vertex}.  We
conclude in Sec.~\ref{sec:conclusions}.  Our conventions and a
generalization of the Wolfenstein parametrization to the case of
complex $\lambda$ and $A$ parameters are given in the~\ref{sec:app}.

\begin{figure}[tb]
\includegraphics[width=0.5\textwidth]{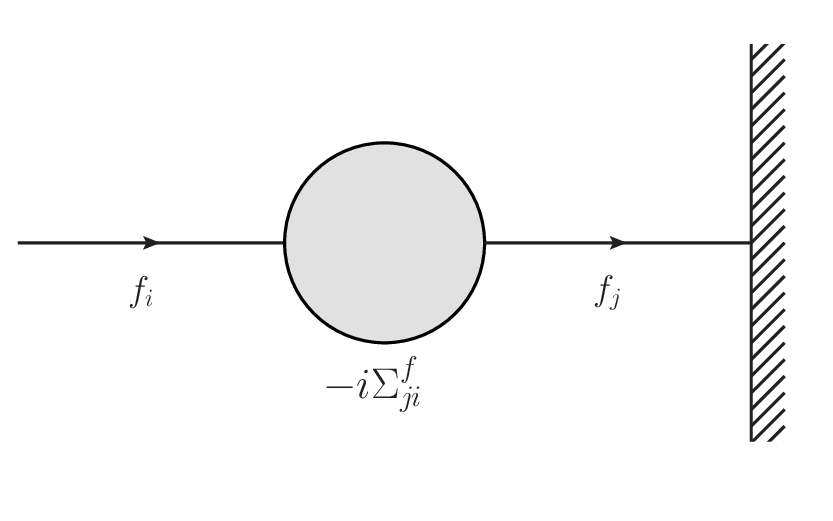}
\caption{Self-energy inducing wave-function rotation in
  flavor-space. 
}
\label{fig:self_energy}
\end{figure}

\section{Chirally-enhanced contributions to self-energies}
\label{sec:Self-energies}

In this section we calculate all chirally-enhanced contributions from
fermion self-energies in the general MSSM.  We first give the complete
formulae and then extract the leading order in $v/\Msusy$, up to which
we will be able to give analytic results for the effective vertices.\medskip

\subsection{General remarks}
In general, it is possible to decompose any self-energy (see
Fig.~\ref{fig:self_energy}) into chirality-flipping and
chirality-conserving parts in the following way (in what follows we
denote the flavor of the incoming (outgoing or ``final'') fermion by
$i$ ($j$ or $f$), respectively):
\begin{equation}
\Sigma_{ji}^f(p) = \left( {\Sigma_{ji}^{f\,LR}(p^2) + \ps\Sigma
  _{ji}^{f\,RR}(p^2) } \right)P_R + \left( {\Sigma_{ji}^{f\,RL}(p^2) +
  \ps\Sigma_{ji}^{f\,LL}(p^2) } \right)P_L\,
\label{self-energy-decomposition}
\end{equation}
Note that the chirality-changing parts $\Sigma_{ji}^{f\,LR}$ and
$\Sigma_{ji}^{f\,RL}$ have mass dimension 1 and are related through
\begin{equation}
   \Sigma_{ji}^{f\,LR}(p^2)=\Sigma_{ij}^{f\,RL*}(p^2),
\label{eq:hermSelf}
\end{equation}
while the hermitian chirality-conserving parts $\Sigma_{ji}^{f\,LL} =
\Sigma_{ij}^{f\,LL\star}$ and $\Sigma_{ji}^{f\,RR} =
\Sigma_{ij}^{f\,RR\star}$ are dimensionless and in general not related
to each other.  Any loop contribution to the fermion self-energy
involving sfermions and gluinos, charginos or neutralinos can be written as
\bea
\Sigma_{ji}^{f\tilde \lambda \,LR}(p^2) &=& \dfrac{-1}{16\pi^2}
\sum\limits_{s = 1}^6 \sum\limits_{I = 1}^N m_{\tilde \lambda_I}
\Gamma_{f_j \tilde f_s }^{\tilde \lambda_I L*} \Gamma_{f_i \tilde f_s
}^{\tilde \lambda_I R} B_0 \left( p^2 ;m_{\tilde \lambda_I }^2,
m_{\tilde f_s }^2 \right)\,, \nonumber\\
\Sigma_{ji}^{f\tilde \lambda \,RL}(p^2) &=& \dfrac{-1}{16\pi^2}
\sum\limits_{s = 1}^6 \sum\limits_{I = 1}^N m_{\tilde \lambda_I }
\Gamma_{f_j \tilde f_s }^{\tilde \lambda_I R*} \Gamma_{f_i \tilde f_s
}^{\tilde \lambda_I L} B_0 \left( p^2 ;m_{\tilde \lambda_I }^2,
m_{\tilde f_s }^2 \right) \,, \nonumber\\
\Sigma_{ji}^{f\tilde \lambda \,LL}(p^2) &=& \dfrac{-1}{16\pi^2}
\sum\limits_{s = 1}^6 \sum\limits_{I = 1}^N \Gamma_{f_j \tilde f_s
}^{\tilde \lambda_I L*} \Gamma_{f_i \tilde f_s }^{\tilde \lambda_I L}
B_1 \left( {p^2 ;m_{\tilde \lambda_k }^2 ,m_{\tilde f_s }^2 } \right)
\,, \nonumber\\
\Sigma_{ji}^{f\tilde \lambda \,RR}(p^2) &=& \dfrac{-1}{16\pi^2}
\sum\limits_{s = 1}^6 \sum\limits_{I = 1}^N \Gamma_{f_j \tilde f_s
}^{\tilde \lambda_I R*} \Gamma_{f_i \tilde f_s }^{\tilde \lambda_I R}
B_1 \left( {p^2 ;m_{\tilde \lambda_I }^2 ,m_{\tilde f_s }^2 } \right)
\,.
\label{MSSM-self-energies}
\eea

Here $\tilde{\lambda}$ stands for the SUSY fermions ($\tilde{g},
\tilde{\chi^0}, \tilde{\chi^\pm}$) and $N$ denotes their corresponding
number (2 for charginos, 4 for neutralinos and 8 for gluinos).  The
coupling coefficients $\Gamma^{\tilde\lambda_I L(R)}_{f_i \tilde f_s }$ and the loop
functions $B_0$ and $B_1$ are defined in \ref{sec:feyrul} and in
\ref{sec:loopint}.  For low-energy decays with $p^2\sim m_f^2\ll
M_{\rm {SUSY}}^2$, it is possible to expand the loop function in the
small parameter $p^2/M_{\rm{SUSY}}^2$:
\bea
B_0 \left( {p^2 ;m_1^2 ,m_2^2 } \right) &=& B_0 \left( {m_1^2 ,m_2^2 }
\right) + p^2 m_2^2 D_0 \left( {m_1^2 ,m_2^2 ,m_2^2 ,m_2^2 } \right) +
\ldots \nonumber\\
B_1 \left( {p^2 ;m_1^2 ,m_2^2 } \right) &=& \dfrac{1}{2}C_2 \left(
{m_1^2 ,m_2^2 ,m_2^2 } \right) + m^2 p^2 E_2 \left( {m_1^2 ,m_2^2
  ,m_2^2 ,m_2^2 ,m_2^2 } \right) + \ldots
\label{eq:bcexpand}
\eea
For most processes, it is sufficient to evaluate the self-energies at
vanishing external momentum.  Further, only the chirality-flipping
part of a self-energy ($\Sigma_{ji}^{f\,LR}, \Sigma_{ji}^{f\,RL}$) can
be enhanced in the MSSM either by a factor $\tan\beta$
\cite{Hall:1993gn} or by a factor $A^{f}_{ij}/(Y^{f}_{ij}\Msusy)$
\cite{Crivellin:2008mq}.  Therefore, we neglect the chirality-conserving 
parts $\Sigma_{ji}^{f\,LL,RR}$ in the following.\medskip

We parametrize the $6\times 6$ sfermion mixing matrices as
\begin{equation}
\mathcal{M}^2_f\,=\,\begin{pmatrix} \Delta^{f\,LL} &
\Delta^{f\,LR} \\ \Delta^{f\,LR\dagger} &
\Delta^{f\,RR}
\end{pmatrix}
\label{eq:Deltas}
\end{equation}
with $\Delta^{f\,XY}$ being $3\times 3$ matrices in flavor-space.  The
numerical values for the $\Delta_{ij}^{f\,XY}$ depend on the chosen
basis for the sfermion fields. It is common to choose for the quark fields
the basis in which the Yukawa couplings are dioagonal and, 
in order to have manifest supersymmetry in the superpotential, to subject 
the squarks to the same rotations as the quarks. The resulting basis
for the super-fields is called super-CKM basis.\medskip

We choose the super-CKM basis for the squark mass matrices by
requiring that the fundamental bare Yukawa couplings $Y^{q\,(0)}$ in
the superpotential are diagonal in flavor space. As discussed in
Ref.~\cite{Crivellin:2009ar}, such a definition of the super-CKM basis
has several advantages compared to an "on-shell" definition in which
the physical quark masses are diagonal instead:
\begin{itemize}
\item The definition of the $\Delta_{ij}^{q\,XY}$ does not depend on
  the renormalization scheme used for the fermion mass matrices
  $m^q_{ij}$.
\item The basis for the $\Delta_{ij}^{q\,XY}$ is defined at the level
  of bare quantities, so that their definition remains valid to all
  orders in perturbation theory.  A choice of basis with the
  renormalized Yukawa couplings $Y^q$ being diagonal, on the other
  hand, requires a redefinition of the $\Delta_{ij}^{q\,XY}$ at every
  order in perturbation theory.
\item In our super-CKM basis the squark mass matrices are diagonal in
  a scenario of flavor-blind SUSY breaking terms. If an on-shell
  definition is used instead, the bare Yukawa couplings $Y^{q(0)}$
  entering the squark mass matrices are not diagonal anymore and the
  squark mass matrices develop flavor off-diagonal entries even in
  case of flavor-blind SUSY breaking terms.
\end{itemize}
The elements 
\begin{eqnarray}
\Delta^{u\,LR}_{ij}&=&-v_u A^u_{ij}\;-\;v_d A^{\prime
  u}_{ij}\;-\;v_d\,\mu\, Y^{u_i{(0)}}\, \delta_{ij}\,,\nn\\
\Delta^{d\,LR}_{ij}&=&-v_d A^d_{ij}\;-\;v_u A^{\prime d}_{ij}\;-\;v_u\,
\mu\, Y^{d_i(0)}\, \delta_{ij}\,,\nn\\ 
\Delta^{\ell\,LR}_{ij}&=&-v_d A^\ell_{ij}\;-\;v_u A^{\prime
  \ell}_{ij}\;-\;v_u\,\mu\, Y^{\ell_i(0)}\, \delta_{ij}\,
\label{DeltaLR}
\end{eqnarray}
and $\Delta^{f\,RL}_{ij}=\Delta^{f\,LR*}_{ji}$ flip the "chiralities".
Appearing in gluino-squark, chargino-sfermion or neutralino-sfermion
contributions to fermion self-energies, they generate
chirality-enhanced effects with respect to the tree-level masses if
they involve the large vev $v_u$ ($\tan\beta$-enhancement for
down-quark/lepton self-energies) or a trilinear $A^{(\prime)f}$-term
($A^{(\prime)f}_{ij}/(Y^f_{ij}\Msusy)$-enhancement).  \medskip

The couplings $\Gamma^{\tilde \lambda_I L(R)}_{f_i \tilde f_s }$ in
\eq{MSSM-self-energies} depend on the corresponding sfermion mixing
matrix and thus on the elements $\Delta^{f\,LR}_{ij}\sim v\Msusy$
entering the sfermion mass matrices.  As non-polynomial functions of
these terms, the $\Gamma^{\tilde\lambda_I L(R)}_{f_i \tilde f_s }$
contain all orders in $(v/\Msusy)^n$ ($n=0,1,2,...$).  However, in the
limit $\Msusy\gg v$ this power series rapidly converges and only the
first terms in the expansion are relevant.  The assumption $\Msusy\gg
v$ is an excellent approximation to the full theory as soon as one
takes into account bounds from direct SUSY searches
\cite{Crivellin:2010er}.  Qualitatively this can be understood as
follows: the off-diagonal mass-insertion terms induce a splitting of
the sfermion masses of the form $m_{\tilde{f}_{1,2}}^2\sim \Msusy^2\pm
v\Msusy$.  Therefore, in order to establish sfermion masses which
respect the lower bounds from direct searches, a hierarchy $\Msusy\gg
v$ is needed to a certain degree.  In practice it is then sufficient
to work to leading order in $v/M_{\rm{SUSY}}$\footnote{Only in the
  case of very light SUSY masses, negative $\mu$ (which is disfavored
  by the anomalous magnetic moment of the muon) and large $\tan\beta$,
  big corrections (compared to the decoupling limit) in the relation
  between the bottom-quark Yukawa coupling and its mass are
  possible.}.  This simplifies the expressions for the self-energies
and will later allow us to give analytic formulae for the effective
vertices.\medskip

The leading terms in the expansion of the self-energies in $v/\Msusy$
do not vanish in the limit of infinitely heavy SUSY masses (if all
dimensionful SUSY parameters are rescaled simultaneously).  We refer
to the approximation in which only such non-decoupling terms are kept
as "the decoupling limit".  Note, however, that even when working only
to leading order in $v/\Msusy$ we do not integrate out the SUSY
particles.  We rather work in the framework of
Refs.~\cite{Carena:1999py, Hofer:2009xb, Crivellin:2009ar,
  Crivellin:2010er} in which the SUSY-particles are kept as dynamical
degrees of freedom and which thus permits a consistent formulation of
effective couplings involving these particles.
\medskip

To leading order in $v/\Msusy$, the chirality-flipping elements
$\Delta^{f\,LR}$ can be neglected in the determination 
of sfermion mixing matrices. The sfermion mass matrices are then
block-diagonal and diagonalized by the mixing matrices $W^f$:
\begin{eqnarray}
W^{d\dagger}\,\mathcal{M}^2_d\,W^d&=&\textrm{diag}\left(m_{\tilde{q}_1^L}^2,m_{\tilde{q}_2^L}^2,
m_{\tilde{q}_3^L}^2,m_{\tilde{d}^R_1}^2,m_{\tilde{d}^R_2}^2,m_{\tilde{d}^R_3}^2\right)\,,\hspace{1cm}
W^d\,=\,\begin{pmatrix} W^{d\,L} & 0 \\ 0 & W^{d\,R} \end{pmatrix}\,,
\nn\\[0.2cm]
W^{u\dagger}\,\mathcal{M}^2_u\,W^u&=&\textrm{diag}\left(m_{\tilde{q}_1^L}^2,m_{\tilde{q}_2^L}^2,
m_{\tilde{q}_3^L}^2,m_{\tilde{u}^R_1}^2,m_{\tilde{u}^R_2}^2,m_{\tilde{u}^R_3}^2\right)\,,\hspace{1cm}
W^u\,=\,\begin{pmatrix} W^{u\,L} & 0 \\ 0 & W^{u\,R} \end{pmatrix}\,,
\nn\\[0.2cm]
W^{\ell\dagger}\,\mathcal{M}^2_\ell\,W^\ell &=&
\textrm{diag}\left(m_{\tilde{\ell}_1^L}^2,m_{\tilde{\ell}_2^L}^2,
m_{\tilde{\ell}_3^L}^2,m_{\tilde{\ell}^R_1}^2,m_{\tilde{\ell}^R_2}^2,m_{\tilde{\ell}^R_3}^2\right)
\,,\hspace{1.2cm} W^\ell\,=\,
\begin{pmatrix} W^{\ell \,L} & 0 \\ 0 &
W^{\ell\, R} \end{pmatrix}\,.\label{eq:Wmat}
\end{eqnarray}
The $3\times 3$\,-\,matrices $W^{f\,L,R}$ ($f=u,d,\ell$) take into
account the flavor mixing in the left- and right-sector of sfermions,
respectively.  Note that $SU(2)_L$-invariance enforces
$\Delta^{u\,LL}\,=\,V^{(0)}\,\Delta^{d\,LL} \,V^{(0)\dagger}$.  Here
$V^{(0)}$ denotes the bare CKM matrix appearing in the diagonalization
of the fundamental Yukawa couplings $Y^{u\,(0)}$, $Y^{d\,(0)}$.  As a
consequence, the masses $m_{\tilde{q}^i_L}$ of left-handed squarks are
the same in the up- and down-sector and the corresponding mixing
matrices are related to each other via the CKM matrix $V^{(0)}$:
\begin{equation}
W^{d\,L}\,=\,W^{q\,L}\,,\hspace{2cm} W^{u\,L}\,=\,V^{(0)}W^{q\,L}\,.
\label{SquarkMixSU2}
\end{equation}
It is further convenient to introduce the abbreviations
\bea
\Lambda_{m\,ij}^{f\,LL} \,=\, W^{f\,L}_{im}\,W^{f\,L\star}_{jm}\,,
&\hskip 1cm & (f=u,d,q,\ell), \nonumber\\ 
\Lambda_{m\,ij}^{f\,RR} \,=\, W^{f\,R}_{im}\,W^{f\,R\star}_{jm}\,,
&\hskip 1cm & (f=u,d,\ell),
\label{eq:Vmat}
\eea
where $i,j,m=1,2,3$ and where index $m$ is not summed over.\medskip

Left-right-mixing of sfermions, on the other hand, is
not described by a mixing matrix but rather treated perturbatively in
the form of two-point $\tilde{f}^R_i$-$\tilde{f}^L_j$ vertices
governed by the couplings $\Delta^{f\,LR}_{ji}$.\medskip 

\subsection{Explicit expressions for the self-energies}

To leading order in $v/\Msusy$, the self-energy with a gluino and a
squark as virtual particles is proportional to one element
$\Delta_{jk}^{q\,LR}$ of the squark mixing matrix (note that the
self-energy scales like $\Delta_{jk}^{q\,LR}/M_{SUSY}$ and thus the
combination is non-decoupling). We have
\begin{eqnarray}
\Sigma_{fi}^{d \tilde{g}\,LR} &=& \dfrac{2\alpha_s}{3\pi}\, m_{\tilde
  g} \sum\limits_{j,k = 1}^3\; \sum\limits_{m,n =
  1}^3 \Lambda_{m\,fj}^{q\,LL}\; \Delta _{jk}^{d\,LR}\;
\Lambda_{n\,ki}^{d\,RR}\; C_0\! \left( m_{\tilde g}^2, m_{\tilde
  q_m^L}^2 ,m_{\tilde d_n^R }^2 \right),\nonumber\\
\Sigma_{fi}^{u \tilde{g}\,LR} &=& \dfrac{2\alpha_s}{{3\pi }}\,
m_{\tilde g} \sum\limits_{j,k,j^\prime,f^\prime = 1}^3\;
\sum\limits_{m,n = 1}^3 V^{(0)}_{ff^\prime}\;\Lambda_{m\,f^\prime
  j^\prime}^{q\,LL} \;V^{(0)\star}_{jj^\prime}\;\Delta_{jk}^{u\,LR}
\; \Lambda_{n\,ki}^{u\,RR}\; C_0\! \left( {m_{\tilde g}^2 ,m_{\tilde
    q_m^L }^2 ,m_{\tilde u_n^R }^2 } \right).
\label{eq:gluinoSE}
 \end{eqnarray}
The matrices $\Lambda_{m\,ki}^{q\,LL,RR}$ $(q=u,d)$ take into account
all powers of chirality-conserving flavor changes induced through the
off-diagonal elements $\Delta_{ij}^{q\,LL,RR}$.  For example
$\Sigma_{11}^{d\,LR}$ also contains a contribution which, in the mass
insertion approximation, would be $\propto \left( \Delta_{13}^{d\,LL}
\; \Delta_{33}^{d\,LR} \; \Delta_{31}^{d\,RR}\right)$.  Therefore,
\eq{eq:gluinoSE} is exact in the decoupling-limit.  The corresponding 
self-energy with flipped chiralities is determined through
\eq{eq:hermSelf}.\medskip

For the neutralino-sfermion contributions to the lepton and quark
self-energies we get
\begin{eqnarray}
\Sigma_{fi}^{\ell\tilde \chi^0 \,LR} &=& \dfrac{1}{16\pi^2}
\left\{\sum\limits_{j,k = 1}^3 \sum\limits_{m,n = 1}^3 g_1^2 M_1\,
\Lambda_{m\,fj}^{\ell\,LL}\, \Delta_{jk}^{\ell\,LR} \,
\Lambda_{n\,ki}^{\ell\,RR} C_0\!  \left( {\left| {M_1 }
  \right|^2,m_{\tilde \ell_{m}^L }^2 ,m_{\tilde \ell_{n}^R }^2 }
\right) \right.  \nonumber\\
&+& \sum\limits_{m = 1}^3 \left[ \frac{1}{\sqrt{2}g_2} M_W
  \sin\beta\;Y^{\ell_i(0)} \Lambda_{m\,fi}^{\ell \,LL} \left( g_2^2
  M_2 \mu\; C_0 \left( \left| M_2 \right|^2,\left| {\mu }
  \right|^2,m_{\tilde \ell_m^L }^2 \right) \right.\right.\nn\\
&-&\left. g_1^2 M_1 \mu C_0 \left( \left| M_1 \right|^2,\left| \mu
        \right|^2,m_{\tilde \ell_m^L }^2 \right) \right) \nonumber\\
&+&\left.  \left.  g_1^2 v_u M_1 \mu\;Y^{\ell_f(0) }
\Lambda_{m\,fi}^{\ell \,RR}\; C_0 \left( {\left| {M_1 }
  \right|^2,\left| {\mu } \right|^2,m_{\tilde \ell_m^R }^2 } \right)
\right] \right\}\,,\nonumber\\
\Sigma_{fi }^{d\tilde \chi^0 \,LR} &=& \dfrac{1}{16\pi^2}\left\{
\sum\limits_{j,k = 1}^3 {\sum\limits_{m,n = 1}^3 -\dfrac{1}{9}g_1^2
  M_1\, {\Lambda_{m\,fj}^{q\,LL}\, \Delta_{jk}^{d\,LR} \,
    \Lambda_{n\,ki}^{d\,RR}}} C_0\! \left( \left| {M_1}
\right|^2,m_{\tilde q_{m}^L }^2 ,m_{\tilde d_{n}^R }^2 \right) \right.
\nonumber\\
&+& \sum\limits_{m=1}^3 \left[ \dfrac{1}{\sqrt{2}g_2}M_W \sin\beta\;
  Y^{d_i(0)} \Lambda_{m\,fi}^{q\,LL}\left( g_2^2 M_2\; \mu\; C_0
  \left( \left| M_2 \right|^2,\left| \mu \right|^2,m_{\tilde q_m^L }^2
  \right) \right. \right.\nonumber\\
&+& \left. \dfrac{g_1^2 }{3} M_1 \mu C_0 \left( \left| M_1 \right|^2,\left|
  \mu \right|^2, m_{\tilde q_m^L}^2 \right) \right) \nonumber\\
&+& \left.  {\left.  \dfrac{1}{3}g_1^2 v_u M_1 \mu\;
    Y^{d_f(0) } \Lambda_{m\,fi}^{d\,RR}\; C_0 \left( {\left| {M_1 }
      \right|^2,\left| {\mu } \right|^2,m_{\tilde d_m^R }^2 } \right)
    \right]} \right\}\,, \nonumber\\
\Sigma_{fi}^{u\tilde \chi^0 \,LR} &=& \dfrac{1}{16\pi^2}\!\!\!
\sum\limits_{j,j^\prime,k = 1}^3\sum\limits_{m,n = 1}^3 \!\dfrac{2}{9}g_1^2 M_1 \;
V^{(0)}_{ff^\prime}\; \Lambda_{m\,f^\prime j^\prime}^{q\,LL}
 V^{(0)\star}_{jj^\prime}\Delta_{jk}^{u\,LR}
\Lambda_{n\,ki}^{u\,RR} C_0\! \left( {\left| {M_1 }
  \right|^2,m_{\tilde q_m^L }^2,m_{\tilde u_n^R }^2 } \right) .
\label{neutralino_SE}
\end{eqnarray}
Finally the chargino-sfermion contributions to the lepton and
down-quark self-energy are given by
\begin{eqnarray}
\Sigma_{fi}^{d\tilde \chi^\pm \,LR} =& -&\dfrac{Y^{d_i(0) }}{16\pi^2}\mu
 \left[\delta_{i3}\,Y^{u_3(0) \star}\sum\limits_{m,n = 1}^3 
  V^{(0)\star}_{3f}\;\Lambda_{m\,33}^{q\,LL}\; V^{(0)}_{33} \;
  \Delta_{33}^{u\,LR\star}\;
  \Lambda_{n\,33}^{u\,RR} \; C_0\! \left( \left| \mu \right|^2
  ,m_{\tilde q_m^L }^2 ,m_{\tilde u_n^R }^2 \right) \right.
  \nonumber\\
&-& \left. g_2^2 v_u  M_2 \sum\limits_{m = 1}^3
  \Lambda_{m\,fi}^{q\,LL} C_0\! \left( {m_{\tilde q_m^L }^2 ,\left|
    \mu \right|^2 ,\left| {M_2 } \right|^2 } \right) \right]\,,
\nonumber \\
\Sigma_{fi }^{\ell\tilde \chi^ \pm \,LR} &=& \dfrac{
  Y^{\ell_i(0) } }{16\pi^2} \mu g_2^2 v_u M_2 \sum\limits_{m
  = 1}^3 \Lambda_{m\,fi}^{\ell \,LL} C_0\! \left( m_{\tilde \ell_m^L
}^2 ,\left| \mu \right|^2 ,\left| {M_2 } \right|^2 \right)\,,
\label{chargino-SE}
\end{eqnarray}
where we have further neglected the small up-type Yukawa couplings of
the first two generations and multiple flavor-changes.  Chargino
contributions to up-quark self-energies cannot be chirally enhanced: a
$\tan\beta$\,-\,enhancement is not possible for up-type self-energies
since the tree-level up-quark masses are not suppressed by $\cos\beta$
(in contrast to the down-quark ones).  An
$A^{(\prime)d}_{ij}/(Y^u_{ij}\Msusy)$\,-\,enhancement, on the other
hand, is neither possible for the third generation, where the large
top Yukawa coupling prevents such an effect, nor for the first two
generations, where the contribution is suppressed by a small down-type
coupling $Y^{d_i}$ ($i=1,2$). Note further that we have neglected
terms proportional to $\cot\beta$ in the chargino- and neutralino mass
matrices\footnote{If one would keep the $\cot\beta$\,-\,suppressed
  terms in the chargino- and neutralino mass matrices, the
  self-energies would be divergent and one would have to go through
  the procedure of infinite renormalization.  In addition, one would
  have to consider also the chirally conserving self-energies
  $\slashed{p}\,\Sigma^{f\,LL,RR}_{ij}(0)$ since they generate, after
  application of the Dirac equation, fermion mass terms of the same
  order in $m_b/\Msusy$ and in $\tan\beta$ as the
  $\cot\beta$-suppressed parts of $\Sigma^{f\,LR}_{ij}$.}. \medskip

We denote the sum of all contributions as
\bea
\Sigma_{fi}^{u\,LR} &=& \Sigma_{fi}^{u\tilde{g}\,LR} +
\Sigma_{fi}^{u\tilde{\chi}^{0}\,LR},\nn\\
\Sigma_{fi}^{d\,LR} &=& \Sigma_{fi}^{d\tilde{g}\,LR} +
\Sigma_{fi}^{d\tilde{\chi}^{0}\,LR} +
\Sigma_{fi}^{d\tilde{\chi}^{\pm}\,LR},\nn\\
\Sigma_{fi}^{\ell\,LR} &=& \Sigma_{fi}^{\ell\tilde{\chi}^{0}\,LR} +
\Sigma_{fi}^{\ell\tilde{\chi}^{\pm}\,LR}.
\eea
In order to simplify the notation it is useful to define the quantity
\begin{equation}
   \sigma^f_{ji}\,=\,\frac{\Sigma_{ji}^{f\,LR}}{\max\{m_{f_j},m_{f_i}\}}\,.
\label{eq:sigdef}
\end{equation}
Here $m_{f_i}$ is the $\overline{\rm{MS}}$ renormalized quark mass
extracted from experiment using the SM prescription. It has to be
evaluated at the same scale as the self-energy $\Sigma_{ji}^{f\,LR}$.
The ratio $\sigma^f_{ji}$ is a measure of the chiral enhancement of
the self-energies with respect to corresponding quark masses.\medskip

For the renormalization of the Yukawa couplings and the CKM matrix it
is important to distinguish between the parts of $\Sigma_{ji}^{f\,LR}$
which contain a Yukawa coupling and/or CKM element and those which do
not.  Furthermore, for the determination of the effective
Higgs-fermion-fermion vertices one has to distinguish between parts of
$\Sigma_{ji}^{f\,LR}$ proportional to different Higgs vev's (we call
terms in $\Sigma_{ji}^{d(u)\,LR}$ proportional to $v_{d(u)}$ to be
``holomorphic'', whereas terms in $\Sigma_{ji}^{d(u)\,LR}$
proportional to $v_{u(d)}$ are called ``non-holomorphic'').  Therefore
we will define several corresponding decompositions of
$\Sigma_{ji}^{f\,LR}$ (or $\sigma^f_{ji}$).\medskip

In the expressions (\ref{eq:gluinoSE})-(\ref{chargino-SE}) each term
in the down-quark (lepton) self-energy $\Sigma_{fi}^{d(\ell)\,LR}$
involves at most one power of the corresponding Yukawa-coupling
$Y^{d(\ell)}$. The up-quark self-energy $\Sigma_{fi}^{u\,LR}$, on the other hand, 
is approximatly independent of $Y^u$ as it is always multiplied by
$\cot\beta$ and can be neglected if one takes into account only
chirally-enhanced contributions.  We make the $Y^{d(\ell)}$-dependence
of the flavor-conserving self-energy $\Sigma_{ii}^{d(\ell)\,LR}$
explicit by decomposing it as
\begin{equation}
\Sigma_{ii}^{d(\ell)\,LR} \;=\;
\Sigma_{ii\,\cancel{Y_i}}^{d(\ell)\,LR} \, + \,
\epsilon_i^{d(\ell)}\,v_u\,\,Y^{d_i(\ell_i)(0)}\,.
\label{eq:epsilon_b}
\end{equation}
In a similar way we decompose the flavor-changing self-energies
$\Sigma_{fi}^{q\,LR}$ ($q=u,d$) with respect to CKM elements.
Concerning the down-type quarks, only $\Sigma^{d\,LR}_{f3}$ ($f=1,2$)
depends on (off-diagonal) CKM elements in the approximation in which
we neglect small mass ratios and multiple flavor-changes.  For
$f\not=i$ we write the enhancement factors $\sigma^d_{fi}$ as
\begin{equation}
\sigma_{fi}^{d} \;=\; \left\{\begin{array}{l}
\widehat{\sigma}^d_{f3}\,+\,\epsilon^d_{\textrm{FC}} V_{3f}^{\left(0
  \right)\star}V_{33}^{(0)}\,,\hspace{0.5cm}\textrm{i=3}\\
\widehat{\sigma}^d_{fi}\,,
\hspace{3.4cm} \textrm{i=1,2} \end{array}\right.\,,
\label{sigmahatd}
\end{equation}
so that the $\widehat{\sigma}^d_{fi}$ do not depend on (off-diagonal)
CKM elements and
\begin{equation}
\varepsilon^d_{FC} = \dfrac{-1}{16\pi^2}\,\mu\,
\dfrac{Y^{d_3(0)}}{m_{d_3}} \sum\limits_{m,n = 1}^{3}
Y^{u_3(0)\star}\,\Lambda_{m\,33}^{q\,LL}\,\Delta_{33}^{u\,LR\star}\,
\Lambda_{n\,33}^{u\,RR}\, C_0 \left( \left| \mu \right|^2 ,m_{\tilde
  q_m^L }^2 ,m_{\tilde u_{n}^R }^2 \right).\label{eq:epsFC}
\end{equation}
For the up-quark self-energy $\Sigma^{u\,LR}_{fi}$ the situation is
more involved.  It depends on the CKM matrix through
$\Lambda^{u\,LL}$, which is related to $\Lambda^{q\,LL}$ via the SU(2)
relation $\Lambda^{u\,LL} = V^{(0)\dagger} \Lambda^{q\,LL} V^{(0)}$ in
the decoupling limit.  Therefore, the bare CKM matrix enters the
gluino- and neutralino-contributions to $\Sigma^{u\,LR}_{fi}$ in
Eqs.~(\ref{eq:gluinoSE}) and (\ref{neutralino_SE}).  However, there
are several reasons why its effect is usually very small. Firstly, a
self-energy diagram with an external top quark cannot be significantly
chirally enhanced as it has to be compared to the large top quark
mass.  Furthermore, effects of the CKM matrix in $\Sigma^{u\,LR}_{fi}$
are proportional to the mass splitting of left-handed squarks (and
cancel completely if the left-handed squark masses are
degenerate\footnote{See Ref.~\cite{ACMD} for a discussion of the
  possibility of non-degenerate squark masses}).  Therefore, in most
cases it is an excellent approximation to assume that the up-quark
self-energies do not depend on (bare) CKM elements and one can set the
CKM elements $V^{(0)}_{ij}$ in Eqs.~(\ref{eq:gluinoSE}) and
(\ref{neutralino_SE}) to their physical values $V_{ij}$. We make this
approximation explicit by writing
\begin{equation}
   \sigma^u_{fi}\,\approx\,\widehat{\sigma}^u_{fi}\label{sigmahatu}
\end{equation}
where $\widehat{\sigma}^u_{fi}$ is understood to be independent of
(off-diagonal) bare CKM elements. For completeness in~\ref{upSE_CKM}
we give analytic expressions for the CKM matrix renormalization which
take into account the dependence of the up-squark sector on the CKM
elements.\medskip

For the discussion of the effective Higgs vertices in
Sec.~\ref{sec:higgs_vertex} we also need a decomposition of
$\Sigma_{ji}^{f\,LR}$ into its holomorphic and non-holomorphic parts,
as mentioned above.  In the decoupling limit all holomorphic
self-energies are proportional to $A$-terms.  Thus we denote the
holomorphic part as $\Sigma_{ji\,A}^{f\,LR}$, while the
non-holomorphic part is denoted as $\Sigma_{ji}^{\prime f \,LR}$.
Then we have
\begin{equation}
\Sigma_{ji}^{f\,LR} = \Sigma_{ji\,A}^{f\,LR} + \Sigma_{ji}^{\prime f
  \,LR}\,,  \label{HoloDeco}
\end{equation}
and the corresponding equation for $\sigma_{ji}^{f\,LR}$.  \medskip

For the decomposition of the self-energies we have assumed that the
$A$-terms and the bilinear soft squark mass terms do not depend on CKM
elements or Yukawa couplings. For example in symmetry-based MFV
\cite{D'Ambrosio:2002ex} this is not the case and those parameters
carry an additional dependence on CKM elements and Yukawa
couplings. Then the self-energies are no longer linear in the Yukawa
couplings and an analytic resummation, as we will perform in the
following chapter, is impossible. In such cases one has to rely on an
iterative procedure in order to determine the bare Yukawa couplings
and bare CKM elements\footnote{Iteration is also needed if the results
  of Ref.~\cite{Ellis:2009di} are applied to the general MSSM because
  in~\cite{Ellis:2009di} the soft SUSY-breaking terms are parametrized
  in terms of Yukawa couplings.}.

\section{Renormalization}
\label{sec:Renormalization}

In this chapter we consider the general effects of the finite
chirally-enhanced self-energies on mass and wave-function
renormalization of fermions and on the renormalization of the CKM
matrix. We do not consider the renormalization of the PMNS matrix
because the renormalization effects are known to be very small
\cite{Girrbach:2009uy,Crivellin:2010gw}.\medskip

\subsection{Renormalization of fermion masses and Yukawa couplings}

Chirally-enhanced self-energies modify the relation between the bare
Yukawa couplings $Y^{f_i(0)}$ and the corresponding physical fermion
masses $m_{f_i}$. In our discussion we concentrate on the quark case
postponing conclusions for the lepton case to the end of this
section. Considering only chirally-enhanced corrections, the physical
quark mass is given by
\begin{equation}
m_{q_i }  \;=\; v_q Y^{q_i(0)}  \,+\, \Sigma_{ii}^{q\,LR},\hspace{1cm} (q=u,d) \label{mq-Yq}.
\end{equation}

\eq{mq-Yq} implicitly determines the bare Yukawa couplings
$Y^{q_i(0)}$ for a given set of SUSY parameters.  The actual values
and physical meaning of the renormalized $Y^{q_i}$ depend, of course,
on the renormalization scheme chosen for $Y^{q_i}$.  Thus, to finite
order in perturbation theory, the Feynman amplitude for a given
process would depend on the chosen scheme.  However, in all-order
resummed expressions the scheme dependence drops out and the final
results only depend on the (finite) bare Yukawa couplings
$Y^{q_i(0)}$, which are scheme independent\footnote{Even though the
  bare Yukawa couplings $Y^{q_i(0)}$ are independent of the
  renormalization scheme applied to $Y^{q_i}$, their values depend on
  the choice of the SUSY input parameters, i.e. on the renormalization
  scheme chosen in the squark sector \cite{Hofer:2009xb}.}.\medskip

The self-energy on the right-hand side of \eq{mq-Yq} can in principle
contain arbitrarily many powers of Yukawa couplings.  Therefore, an
analytic solution of \eq{mq-Yq} for $Y^{q_i(0)}$ is not possible in
the general case.  However, since the terms in $\Sigma_{ii}^{q\,LR}$
with higher powers of $Y^{q_i(0)}$ are suppressed by higher powers of
$v/\Msusy$, a numerical solution of \eq{mq-Yq} can be easily achieved
using iterative methods.  For a detailed description of such procedure
in the MFV case we refer to Ref.~\cite{Hofer:2009xb}.  It is obviously
still useful to have an approximate analytic formula at hand, and we
derive it using the decoupling limit.
\medskip

In the up-quark sector the enhanced terms in the self-energy
$\Sigma_{ii}^{u\,LR}$ are independent of $Y^{u_i(0)}$. Therefore
\eq{mq-Yq} is easily solved for $Y^{u_i(0)}$ and one finds
\begin{equation}
Y^{u_i(0)} = \left(m_{u_i } - \Sigma_{ii}^{u\,LR}\right)/v_u .
\label{mu-Yu}
\end{equation}\medskip

In the down-quark sector, if we restrict ourselves to the decoupling
limit where we have terms proportional to one power of $Y^{d_i(0)}$ at
most, we recover the well-known resummation formula for
$\tan\beta$-enhanced corrections, with an extra correction due to the
$A$-terms. The resummation formula is given by
\begin{equation}
Y^{d_i(0)} = \dfrac{m_{d_i} - \Sigma_{ii\,\cancel{Y_i}}^{d\,LR}}{v_d
  \left( {1 + \tan\beta \varepsilon_i^d } \right)}
\label{md-Yd}
\end{equation}
with $\epsilon_i^d$ and $\Sigma_{ii\,\cancel{Y_i}}^{d\,LR}$ defined
through \eq{eq:epsilon_b}.\medskip

Finally, we note that all statements of this section concerning
down-quarks can directly be transferred to the lepton sector.  In
particular the Yukawa coupling $Y^{\ell_i(0)}$ is obtained from
\eq{md-Yd} by replacing fermion index $d$ for $\ell$, except for the
vev.  \medskip

\subsection{Fermion wave-function renormalization}
\label{sec:FCself}

The flavor-changing self-energies $\Sigma^{f\,LR}_{fi}$ induce
wave-function rotations
\begin{equation}
   \psi^{f\,L}_{i}\,\longrightarrow\, U^{f\,L}_{ij}\,\psi^{f\,L}_j\,,\hspace{2cm}
   \psi^{f\,R}_{i}\,\longrightarrow\, U^{f\,R}_{ij}\,\psi^{f\,R}_j
\end{equation}
in flavor-space which have to be applied to all external fermion
fields. We decompose $U^{f\,L,R}_{ij}$ as
\begin{equation}
U^{L,R}_{ij}=\delta_{ij}+\Delta U^{q\,L,R\,(1)}_{ij}+\Delta
U^{q\,L,R\,(2)}_{ij}+...
\end{equation}
where the superscripts denote the respective loop order. 
At the one-loop level $\Delta U^{q\,L}_{fi}$ is given by
\cite{Crivellin:2008mq}
\begin{equation} \renewcommand{\arraystretch}{2.5}
\Delta U^{f\,L\,(1)}  \,=\, 
\left( {\begin{array}{*{20}c}
0 & \sigma^f_{12}\,+\,\dfrac{m_{f_1}}{m_{f_2}}\,\sigma^{f\star}_{21} &
    \sigma^f_{13}\,+\,\dfrac{m_{f_1}}{m_{f_3}}\,\sigma^{f\star}_{31} \\
    -\sigma^{f\star}_{12}\,-\,\dfrac{m_{f_1}}{m_{f_2}}\,\sigma^{f}_{21} & 0 &
    \sigma^f_{23}\,+\,\dfrac{m_{f_2}}{m_{f_3}}\,\sigma^{f\star}_{32} \\
    -\sigma^{f\star}_{13}\,-\,\dfrac{m_{f_1}}{m_{f_3}}\,\sigma^{f}_{31} &
    -\sigma^{f\star}_{23}\,-\,\dfrac{m_{f_2}}{m_{f_3}}\,\sigma^{f}_{32} & 0 \\
\end{array}} \right),
\label{deltaU1}
\end{equation}
where we have neglected terms which are quadratic or of higher order
in small quark mass ratios.  However, for transitions between the
third and the first generation also two-loop corrections are important
\cite{Crivellin:2008mq,Crivellin:2010gw}.  They read
\begin{equation}\renewcommand{\arraystretch}{2}
\Delta U^{f\,L\,\left( 2 \right)}  = \left( {\begin{array}{*{20}c}
- \dfrac{1}{2}\,\left|\sigma^f_{12}\right|^2\,-\,\dfrac{1}{2}\,
\left|\sigma^f_{13}\right|^2 &
-\dfrac{m_{f_3}}{m_{f_2}}\,\sigma^f_{13}\,\sigma^f_{32} &
\dfrac{m_{f_2}}{m_{f_3}}\,\sigma^f_{12}\,\sigma^{f\star}_{32} \\
\dfrac{m_{f_3}}{m_{f_2}}\,\sigma^{f\star}_{13}\,\sigma^{f\star}_{32} &
- \dfrac{1}{2} \,\left|\sigma^f_{12} \right|^2
\,-\,\dfrac{1}{2}\left|\sigma^f_{23}\right|^2 &
\dfrac{m_{f_2}}{m_{f_3}}\,\sigma_{21}^f\,\sigma^{f\star}_{31} \\
\sigma^{f\star}_{12}\,\sigma^{f\star}_{23} &
-\sigma^f_{12}\,\sigma^{f\star}_{13} & -
\dfrac{1}{2}\,\left|\sigma^f_{13}\right|^2\,-\,
\dfrac{1}{2}\,\left|\sigma^f_{23}\right|^2
\\
\end{array}} \right).
\label{WFR2}
\end{equation}
Here only the leading order in the expansion in small quark mass
ratios has been taken into account.  Respecting naturalness
constraints for the CKM hierarchy, only the $3-1$ element in \eq{WFR2}
can be numerically important.  To leading order in the quark mass
ratios the full $U^{f\,L}$ then reads
\begin{equation}
\renewcommand{\arraystretch}{1.5}
U^{f\,L}  = \left( {\begin{array}{*{20}c}
1 & \sigma^f_{12} & \sigma^f_{13} \\
-\sigma^{f\star}_{12} & 1 & \sigma^f_{23} \\
-\left(\sigma^{f\star}_{13}\,-\,\sigma^{f\star}_{12}\,\sigma^{f\star}_{23}\right) & 
-\sigma^{f\star}_{23} & 1 \\
\end{array}} \right).
\label{DeltaU}
\end{equation}
The corresponding expressions for $U^{f\,R}$ are obtained from the
ones for $U^{f\,L}$ by replacing $\sigma^f_{ji} \to
\sigma^{f\star}_{ij}$.

\subsection{Renormalization of the CKM matrix}
\label{subsec:CKMreno}

Application of the rotations in \eq{DeltaU} to the $u_i d_j
W^+$\,-\,vertex renormalizes the CKM elements $V_{ij}$.  The bare CKM
matrix $V^{(0)}$ (stemming from the misalignment between the Yukawa
matrices $Y^{u(0)}$ and $Y^{d(0)}$) can be calculated in terms of the
physical CKM matrix $V$ as
\begin{equation}
V^{(0)}  = U^{u\,L}\, V\, U^{d\,L\dag}.  
\label{CKM-0-ren}
\end{equation}
In the absence of large unnatural cancellations, the rotations
$U^{u\,L}$ and $U^{d\,L}$ preserve the hierarchy of $V$ so that
$V^{(0)}$ has the same hierarchy as $V$.  However, the conventional
Wolfenstein parametrization is not sufficient to describe $U^{u\,L}$,
$U^{d\,L}$ and $V^{(0)}$ since these matrices can have additional
complex phases compared to the physical CKM matrix $V$ (in the case of
$V$ such phases are absorbed by proper redefinition of the quark
fields).  Therefore, we extend the classic Wolfenstein parametrization
in~\ref{sec:wolf}.  In terms of our generalized Wolfenstein
parametrization, defined in \eq{generalized-Wolfenstein}, we have
\bea
V = V\left(\lambda,\,\lambda^2A,\,\lambda^3A(\rho-i\eta),\, 0\right)
\,\equiv\, V\left(v_{12},\,v_{23},\,v_{13},\, 0\right) 
\eea
and
\bea
U^{q\,L} &=& U^{q\,L}\left(\sigma^q_{12},\, \sigma^q_{23},\,
\sigma^q_{13},\,0\right)\hspace{1.5cm}(q=u,d).
\eea
We parametrize $V^{(0)}$ accordingly as
\begin{equation}
V^{(0)}\,=\,\left(v_{12}^{(0)},\,v_{23}^{(0)},\,v_{13}^{(0)},\,v_{\rm{Im}}^{(0)}\right).
\end{equation}\medskip

Using the approximation (\ref{sigmahatu}), the rotation matrix
$U^{u\,L}$ is independent of $V^{(0)}$. We make this explicit by
writing
\begin{equation}
U^{u\,L}\,=\,\widehat{U}^{u\,L}\,=\,\widehat{U}^{u\,L}\left(\widehat{\sigma}^u_{12},
\,\widehat{\sigma}^u_{23},\,\widehat{\sigma}^u_{13},\,0\right).
\end{equation}
The matrix $U^{d\,L}$, on the other hand, consists of a CKM-dependent
and a CKM-independent part since the $\sigma^d_{ji}$ entering
\eq{DeltaU} decompose according to \eq{sigmahatd}.  We transfer this
decomposition to $U^{d\,L}$ writing (in what follows we neglect terms
${\cal O}(\lambda^4)$ and higher)
\begin{equation}
   U^{d\,L}\,=\,U^{d\,L}_{\CKM}\,\widehat{U}^{d\,L}\,.\label{Udeco}
\end{equation}
The CKM-independent part $\widehat{U}^{d\,L}$ is defined by replacing
$\sigma_{ji}^d\to\widehat{\sigma}^d_{ji}$ in \eq{DeltaU}, what amounts
to the generalized Wolfenstein parametrization
\begin{equation}
\widehat{U}^{d\,L}\,=\,\widehat{U}^{d\,L}\left(\widehat{\sigma}^d_{12},
\,\widehat{\sigma}^d_{23},\, \widehat{\sigma}^d_{13},\,0\right).
\end{equation} 
The CKM-dependent part $U^{d\,L}_{\CKM}$ is then given by
\begin{eqnarray}
   \renewcommand{\arraystretch}{1.5}
   U^{d\,L}_{\CKM}&=&U^{d\,L}\,\widehat{U}^{d\,L\dag}\,=\,\begin{pmatrix}
     1 & 0 & V_{31}^{(0)\star}V_{33}^{(0)}\varepsilon_{FC}^d \\
     0 & 1 & V_{32}^{(0)\star}V_{33}^{(0)}\varepsilon_{FC}^d \\
     -V_{31}^{(0)}V_{33}^{(0)\star}\varepsilon_{FC}^{d\star} &
     -V_{32}^{(0)}V_{33}^{(0)\star}\varepsilon_{FC}^{d\star} & 1
   \end{pmatrix}\nonumber\\[0.5cm]
     &=&U^{d\,L}_{\CKM}\left(0,\,V_{32}^{(0)\star}V_{33}^{(0)}\varepsilon_{FC}^d,\,
                    V_{31}^{(0)\star}V_{33}^{(0)}\varepsilon_{FC}^d,\,0\right)\,.
\end{eqnarray}
Inserting the decomposition (\ref{Udeco}) into \eq{CKM-0-ren} we
obtain
\begin{equation}
V^{(0)} \,=\, \left(\widehat{U}^{u\,L}\, V\,
\widehat{U}^{d\,L\dag}\right)\,U^{d\,L\dag}_{\CKM}\,.
   \label{CKM-reno2}   
\end{equation}\medskip
In order to determine $V^{(0)}$, we have to solve \eq{CKM-reno2}. The
right-hand side implicitly depends on $V^{(0)}$ through
$U^{d\,L}_{\CKM}$. As a first step we solve \eq{CKM-reno2} in the
special case $U^{d\,L}_{\CKM}\equiv 1$, i.e.  in the absence of the
contributions governed by $\varepsilon_{FC}^d$.  This means we
calculate
\begin{equation}
   \widetilde{V}  \,=\, \widehat{U}^{u\,L}\, V\, \widehat{U}^{d\,L\dag}\,.
   \label{VtildeDef}
\end{equation}
Exploiting the multiplication rule (\ref{multiplication-rule}) for
generalized Wolfenstein matrices we get
\begin{equation}
   \widetilde{V}\,=\,\left(\widetilde{v}_{12},\,\widetilde{v}_{23},\,
   \widetilde{v}_{13},\,\widetilde{v}_{\rm{Im}}\right)
\end{equation}
with
\begin{eqnarray}
   \widetilde{v}_{12}  &=& v_{12}  \,+\, \widehat{\sigma}_{12}^{u} \,-\, 
                                \widehat{\sigma}^d_{12} \,,\hspace{1.5cm} 
   \widetilde{v}_{23}  \,=\, v_{23} \,+\, \widehat{\sigma}_{23}^{u} \,-\, 
                                \widehat{\sigma}^d_{23} \,,\nonumber\\[0.2cm]
 \widetilde{v}_{13}  &=& v_{13}  \,+\, \widehat{\sigma}_{13}^{u} \,-\, \widehat{\sigma}_{13}^{d}
       \,+\, \widehat{\sigma}_{12}^{u} v_{23}  \,+\, \left( \widehat{\sigma}_{12}^{d} \,-\, 
        \widehat{\sigma}_{12}^{u} \right)
      \widehat{\sigma}_{23}^{d} \,-\, v_{12} \widehat{\sigma}_{23}^{d}\,,\nonumber\\[0.2cm]
 \widetilde{v}_{\rm{Im}}  &=&  v_{12}\, 
{\mathop{\rm Im}\nolimits} \left[\widehat{\sigma}_{12}^{u}\, +\, 
 \widehat{\sigma}_{12}^{d}\right]\,-\,\textrm{Im}\left[\widehat{\sigma}^u_{12}\widehat{\sigma}^{d\star}_{12} 
 \right]\,.
 \label{Vtilde2}
\end{eqnarray}
Switching on $U^{d\,L}_{\CKM}\not= 1$ in a second step and solving
\eq{CKM-reno2} for $V^{(0)}$, we finally find
\begin{equation}
V^{(0)}=V^{(0)}\left(\widetilde{v}_{12},\,\frac{\widetilde{v}_{23}}{1
  - \varepsilon^d_{FC}}, \,\frac{\widetilde{v}_{13}}{1 -
  \varepsilon^d_{FC}},\,\widetilde{v}_{\rm{Im}}\right).
\end{equation}
Explicitly written down, this matrix reads
\begin{equation}
\renewcommand{\arraystretch}{1.5}
V_{}^{\left( 0 \right)}  = \left( {\begin{array}{*{20}c}
   {1 - \dfrac{{\left| {\widetilde{v}_{12} } \right|^2}}{2} + 
   i\,\widetilde{v}_{\rm{Im}} } & {\widetilde{v}_{12} } & 
   {\dfrac{{\widetilde{v}_{13} }}{{1 - \varepsilon_{FC}^d }}}  \\
   { - \widetilde{v}_{12}^{\star} } & {1 - \dfrac{{\left| {\widetilde{v}_{12} } \right|^2}}{2} 
   - i\,\widetilde{v}_{\rm{Im}} } & {\dfrac{{\widetilde{v}_{23} }}{{1 - \varepsilon_{FC}^d }}}  \\
 \dfrac{{\widetilde{v}_{12}^{\star} \widetilde{v}_{23}^{\star} - \widetilde{v}_{13}^{\star} }}{{1 - \varepsilon_{FC}^{d\star} }} & { - \dfrac{{\widetilde{v}_{23}^{\star} }}{{1 - \varepsilon_{FC}^{d\star} }}} & 1  \\
\end{array}} \right)\,.\label{barCKM}
\end{equation}\medskip

We see that the elements $\widetilde{v}_{13}$ and $\widetilde{v}_{23}$
in \eq{barCKM} are scaled by a factor $1/(1 - \varepsilon_{FC}^d)$.
This generalizes the observation of ref.~\cite{Hofer:2009xb}, where it
has been found that the Wolfenstein parameter $A$ is scaled by this
factor in the MFV version of the MSSM, to the case of general flavor
violation.\medskip

\subsection{Proper renormalization sequence}
\label{subsec:renseq}

The determination of the bare Yukawa couplings and bare CKM matrix is
complicated by the fact that the corresponding equations, defined in
the previous sections, are entangled.  We give here a detailed recipe
on how to determine these quantities step by step.\medskip

\noindent 1. One should start from the calculation of the bare Yukawa
couplings.
\begin{itemize}
\item[a)] Calculate the flavor-conserving self-energies
  $\Sigma^{u\,LR}_{ii}$ in the up-sector from \eq{eq:gluinoSE} and
  \eq{neutralino_SE}.  Note that $\Sigma^{u\,LR}_{ii}$ is independent
  of any Yukawa coupling since we neglect terms proportional to
  $\cot\beta$.  Determine the bare up-quark Yukawa couplings
  $Y^{u_i(0)}$ via \eq{mu-Yu}.
\item[b)] Having the bare top-quark Yukawa coupling at hand, extract
  the chargino contribution to $\epsilon^{d}_3$ (defined in
  \eq{eq:epsilon_b}) from \eq{chargino-SE}.  Calculate also all other
  contributions to $\epsilon^d_3$ as well as to the
  $Y^{d_3(0)}$-independent part $\Sigma^{d\,LR}_{33\,\cancel{Y^d_3}}$
  of the flavor-conserving self-energy $\Sigma^{d\,LR}_{33}$ from
  Eqs.~(\ref{eq:gluinoSE})\,-\,(\ref{chargino-SE}).  For this step one
  can neglect small contributions proportional to the strange- or
  down-Yukawa couplings, which are still undetermined.  Calculate the
  bare bottom Yukawa coupling $Y^{d_3(0)}$ from \eq{md-Yd}.
\item[c)] Calculate $\epsilon^d_2$ and
  $\Sigma^{d\,LR}_{22\,\cancel{Y^d_2}}$ for the strange quark
  analogously to step 1b) for the bottom quark.  In the calculation,
  $Y^{d_3(0)}$ should be set to the value determined in step 1b),
  while $Y^{d_1(0)}$ can again be neglected.  Compute $Y^{d_2(0)}$
  according to \eq{md-Yd}.
\item[d)] Proceed in the same way for $Y^{d_1(0)}$ using the already
  determined values for $Y^{d_2(0)},Y^{d_3(0)}$.
\end{itemize}
\medskip

\noindent 2. In the next step, the bare CKM matrix and the field rotation
matrices can be determined.
\begin{itemize}
\item[a)] Use the value of the bare Yukawa couplings to determine
  $\epsilon_{FC}^d$ and the CKM-independent self-energy parameters
  $\widehat{\sigma}^u_{ij}$ and $\widehat{\sigma}^{d}_{ij}$ from
  Eqs.~(\ref{eq:gluinoSE})\,-\,(\ref{chargino-SE}) according to the
  decompositions (\ref{sigmahatd}),(\ref{sigmahatu}).  This allows to
  compute the bare CKM matrix $V^{(0)}_{ij}$ with help of
  Eqs.~(\ref{Vtilde2}) and (\ref{barCKM}).
\item[b)] Next, insert the bare Yukawa couplings and $V^{(0)}_{ij}$
  into \eq{eq:sigdef} in order to compute the full $\sigma^d_{ij}$.
  Also $\sigma^u_{ij}$ should be recalculated using the bare
  $V^{(0)}_{ij}$ (instead of the $V_{ij}$ which have been used in the
  calculation of the $\widehat{\sigma}^u_{ij}$). With the
  $\sigma^u_{ij}$ and $\sigma^d_{ij}$ at hand, one calculates the
  rotation matrices $U^{q\,L,R}_{ij}$ ($q=u,d$) in \eq{DeltaU}.
\end{itemize}
\bigskip

The procedure used for the down quarks applies to the
charged leptons as well.\medskip

\section{Effective fermion vertices}
\label{sec:effvert}

Having determined in the previous section the bare Yukawa couplings
and the bare CKM matrix we are now in a position to calculate the
effective gaugino(higgsino)-fermion-sfermion and the effective
Higgs-fermion-fermion vertices in the general MSSM.\medskip

\subsection{Effective gaugino-fermion-sfermion and
  higgsino-fermion-sfermion vertices}
\label{sec:effective-vertices}

In order to calculate the effective gaugino(higgsino)-fermion-sfermion
vertices, one has to take the Feynman-rules given in~\ref{sec:feyrul}
and substitute in the couplings $\Gamma_{f_j \tilde f_s }^{\tilde
  \lambda\, L,R}$ the tree-level Yukawa couplings and the CKM matrix
by the corresponding bare quantities (since the Feynman-rules
in~\ref{sec:feyrul} go beyond the decoupling limit approximation, one
should also recalculate the sfermion masses and sfermion mixing
matrices with the use of bare quantities). In addition, one has to
apply the wave-function rotations to the fermion fields replacing
$\Gamma_{f_j\tilde f_s }^{\tilde \lambda\, L,R}$ by
\begin{equation}
\Gamma_{f_i \tilde f_s
}^{\tilde\lambda\,L\,\rm{eff}}\,=\,U^{f\,L}_{ji}\,\Gamma_{f_j \tilde
  f_s }^{\tilde \lambda\, L},\hspace{2cm} \Gamma_{f_i \tilde f_s
}^{\tilde \lambda\,R\,\rm{eff}}\,=\,U^{f\,R}_{ji}\,\Gamma_{f_j \tilde
  f_s }^{\tilde \lambda\, R}\,.\label{effVert}
\end{equation}
If the momentum $p$ flowing through the fermion line satisfies
$p^2\ll\Msusy^2$, the rotations $U^{f\,L,R}_{ji}$ take into account
the effects of flavor-changing chirally-enhanced self-energy
corrections (to leading order in $p^2/\Msusy^2$). For
$p^2\sim\Msusy^2$, on the other hand, no chiral enhancement occurs and
the rotations $U^{f\,L,R}_{ji}$ drop out from internal fermion
lines. Therefore our effective vertices can be applied irrespective of
the momentum flowing through the fermion line.\medskip

The appearance of the rotations $U^{f\,L,R}_{ji}$ in
gaugino(higgsino)-fermion-sfermion vertices is a consequence of the
fact that our super-CKM basis is defined at the level of the bare
Yukawa couplings $Y^{q(0)}$. Therefore, it is natural to ask whether
these effects can be absorbed into the definition of the squark mass
terms if an on-shell definition for the super-CKM basis is used.
Note, however, that at least for the higgsino-parts of the chargino-
and neutralino-vertices, this is impossible: if an on-shell definition
for the super-CKM basis is used, the bare Yukawa couplings
$Y^{q(0)}_{ij}$ develop off-diagonal entries which are related to the
rotation matrices $U^{f\,L,R}_{ji}$. In this way the physical effects
of these rotations would reappear in the higgsino-fermion-sfermion
coupling.  Note further that an absorption of the effects in
gaugino-fermion-sfermion vertices, is only possible as long as the
bilinear SUSY breaking terms are independent free parameters. As soon
as a structure resulting from a SUSY breaking mechanism (like
gravity-mediation or gauge-mediation) is assumed for them, an
arbitrary redefinition is not possible anymore and the effects of
$U^{f\,L,R}_{ji}$ become physical here as well.\medskip

\subsection{Effective Higgs-fermion-fermion vertices}
\label{sec:higgs_vertex}

\begin{nfigure}{t}
\includegraphics[width=0.8\textwidth]{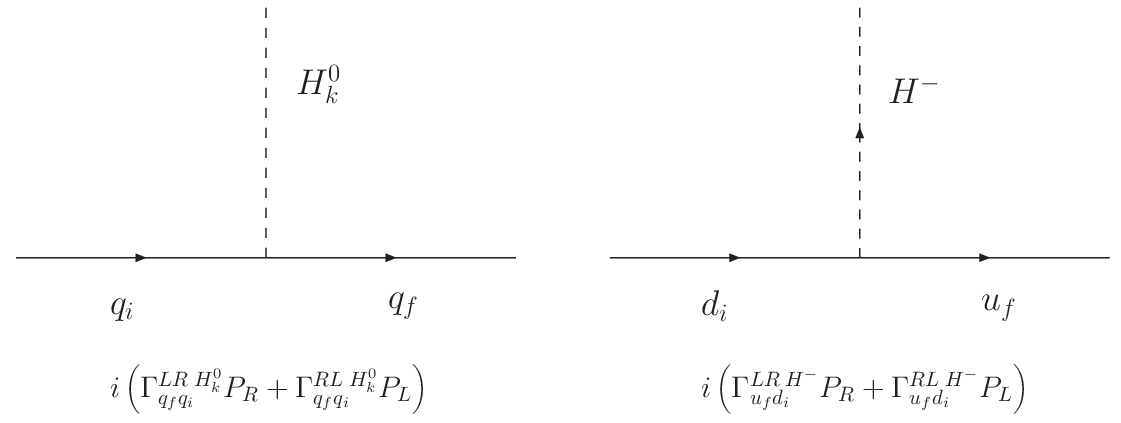}
\caption{Higgs-quark vertices with the corresponding Feynman-rules.}
\label{fig:Higgs-Quark-Coupling}
\end{nfigure}

Also Higgs-fermion-fermion couplings receive chirally-enhanced
corrections from the Yukawa- and CKM-renormalization and from the
fermion wave-function rotations. In addition, we face a new class of
chirally-enhanced effects: the Higgs coupling itself involves a Yukawa
coupling $Y^f$ with $Y^f\ll 1$ for $f\not=t$.  Therefore a genuine
vertex correction which avoids the $Y^f$\,-\,suppression by coupling
to the Higgs via the $A^{(\prime)f}$\,-\,term can be chirally enhanced
with respect to the tree-level vertex. The loop suppression can be
alleviated by a factor $A^{(\prime)f}_{ij}/(Y^f_{ij}\Msusy)$ in this
case. Note that this type of chiral enhancement cannot replicate
itself at higher orders in perturbation theory, so that no resummation
is needed.\medskip

Since all corrections to gaugino(higgsino)-fermion-sfermion vertices
were due to fermion self-energies, they did not depend on the momenta
of the SUSY particles but only on the momentum $p$ of the fermion.  As
shown in Refs.~\cite{Carena:1999py,Hofer:2009xb}, chirally-enhanced
effects only occur for $p^2\ll M_{\rm{SUSY}}^2$. Therefore, such
effects are local and can be cast into effective Feynman rules without
any further assumptions.  In the case of the genuine vertex
corrections to the Higgs-fermion-fermion couplings, the situation is
different.  These corrections are chirally enhanced, independently of
the scale of the external momenta.  In order to derive effective
Feynman rules for these vertices, however, we have to assume that the
external momenta are much smaller than the masses of the virtual SUSY
particles running in the loop. This assumption limits the
applicability of the resulting Feynman rules: if
$m_{H^0},m_{A^0},m_{H^{\pm}}\ll \Msusy$ ($H^0,A^0,H^{\pm}$ denote the
neutral CP-even, CP-odd and the charged Higgs boson, respectively),
they can be used for all processes including diagrams where the Higgs
bosons are involved in a loop. If this hierarchy is not satisfied,
they can only be used for processes in which the momentum-flow through
the Higgs-fermion-fermion vertex is small compared to $\Msusy$.
Important examples for processes of the latter kind are the Higgs
penguins contributing to $B_{d,s}\to\mu^+\mu^-$, $B^+\to\tau^+\nu$ or
the double Higgs penguin contributing to $\Delta F=2$
processes.\medskip

Effective Higgs-fermion-fermion vertices have been calculated in
Ref.~\cite{Crivellin:2010er}, but only the gluino-squark contributions
have been taken into account.  We extend the results of
Ref.~\cite{Crivellin:2010er} by including also chargino-fermion and
neutralino-fermion corrections.  In Ref.~\cite{Crivellin:2010er} two
different derivations of the effective Higgs vertices have been
presented: the first one, using a diagrammatic method, delivers a
result valid to all orders in $v/\Msusy$, while the second one, using
an effective theory approach, reproduces only the leading order in
$v/\Msusy$. It turned out that the leading order in $v/\Msusy$ is an
excellent approximation to the full approach and there is no reason
why this statement should not be true for the chargino and neutralino
contributions. Furthermore, since we restricted ourselves to leading
order in $v/\Msusy$ in the resummation of the Yukawa couplings (which
enter the Higgs coupling), for consistency we should rely on this
approximation in calculating the genuine vertex corrections as well.
Therefore, we will use the effective field theory approach in our
study of the Higgs-fermion-fermion couplings which simplifies the
calculations.  This means that in contrast to the previous sections we
really integrate out the SUSY particles and remove them as dynamical
degrees of freedom, limiting somewhat the applicability of the
effective Higgs vertices as discussed in the previous
paragraph.  \medskip

The resulting effective Yukawa-Lagrangian is that of a general 2HDM
and we parametrize it (in the super-CKM basis) as
\bea 
\mathcal{L}^{eff}_Y &=& \bar{Q}^a_{f\,L} \left[
  (Y^{d_i(0)}\delta_{fi}+E^d_{fi}) \epsilon_{ab}H^{b\star}_d\,-\,E^{\prime
    d}_{fi} H^{a}_u \right]d_{i\,R}\nn\\
&-&\bar{Q}^a_{f\,L} \left[ (Y^{u_i(0)} \delta_{fi} +
  E^u_{fi})\epsilon_{ab}H^{b\star}_u \,+\, E^{\prime u}_{fi} H^{a}_d
  \right]u_{i\,R}\,+\,h.c.
\eea
Here $a$, $b$ denote $SU(2)_L$\,-\,indices and $\epsilon_{ab}$ is the
two-dimensional antisymmetric tensor with $\epsilon_{12}=1$.  Apart
from the Yukawa-couplings $Y^{u_i}$ and $Y^{d_i}$, we have in the
effective theory loop-induced holomorphic couplings $E^q_{fi}$ and
non-holomorphic couplings $E^{\prime q}_{fi}$ ($q=u,d$).  In the
general MSSM these couplings can be expressed in terms of the
corresponding self-energies, which also decompose into a holomorphic
and a non-holomorphic part according to \eq{HoloDeco}.  We have
\begin{equation}
   E^d_{ij}\,=\,\dfrac{\Sigma^{d\,LR}_{ij\,A}}{v_d}\,,\hspace{1.5cm} 
   E^{\prime d}_{ij}\,=\,\dfrac{\Sigma^{\prime\, d\,LR}_{ij}}{v_u}\,\hspace{1.5cm}
   E^u_{ij}\,=\,\dfrac{\Sigma^{u\,LR}_{ij\,A}}{v_u}\,,\hspace{1.5cm} 
   E^{\prime u}_{ij}\,=\,\dfrac{\Sigma^{\prime\, u\,LR}_{ij}}{v_d}\,.
\end{equation}
These effective couplings are in principle loop-suppressed compared to
the tree-level $Y^{d_i(0)}$, $Y^{u_i(0)}$ but a chiral
enhancement of $A^{(\prime)q}_{ij}/(Y^{q}_{ij}\Msusy)$
can compensate for this suppresion.\medskip

In our effective theory approach, the wave-function
rotations\footnote{Note that even though these rotations are identical
  to the ones in \eq{DeltaU} their origin is different in the
  effective field theory approach. The matrices $U^{q\,L,R}$ are now
  obtained by a perturbative diagonalization of the (physical) quark
  mass matrices (see Ref.~\cite{Crivellin:2010er} for details).}
modify the effective Lagrangian as follows~\cite{Crivellin:2010er}:
\bea
\mathcal{L}_Y^{eff} = &-& \bar d_{f\,L} \left[\left(\dfrac{{m_{d_i }
  }}{{v_d }}\delta_{fi} - \widetilde E_{fi}^{\prime d}\tan\beta
  \right)H_d^{0\star}\,+\,\widetilde E_{fi}^{\prime d}\,H_u^0 \right]d_{i\,R}
\nn \\
&-& \bar u_{f\,L} \left[\left(\dfrac{{m_{u_i } }}{{v_u }}\delta_{fi} -
  \widetilde E_{fi}^{\prime u}\cot\beta \right)H_u^{0\star}\,+\,\widetilde
  E_{fi}^{\prime u}\,H_d^{0} \right] u_{i\,R} \nn\\
&+& \bar u_{f\,L} V_{fj} \left[ {\dfrac{{m_{d_i } }}{{v_d
    }}\delta_{ji}-\left( {\cot \beta + \tan \beta }
    \right)\widetilde E_{ji}^{\prime d}  } \right]\sin \beta H^ + d_{i\,R} \nn \\
&+& \bar d_{f\,L} V_{jf}^{\star} \left[ { \dfrac{{m_{u_i
    } }}{{v_u }}\delta_{ji}-\left( {\tan \beta +
      \cot\beta } \right)\widetilde E_{ji}^{\prime u}  } \right]\cos\beta H^ - u_{i\,R}\,+\,h.c. 
\label{L-Y-FCNC}
\eea
with
\begin{eqnarray}
\widetilde{E}^{\prime q}_{fi}&=&U_{jf}^{q\,L*} E^{\prime q}_{jk}
U_{ki}^{q\,R} \,\approx\, E_{fi}^{\prime q}\,-\, \Delta E_{fi}^{\prime
  q}\,,\nonumber\\[0.3cm]
\Delta E^{\prime q} &=& \left( {\begin{array}{*{20}c} 0 &
    \sigma^q_{12} E_{22}^{\prime q}
    & \hspace{0.5cm}\left(\sigma^q_{13}-\sigma^q_{12}\sigma^q_{23}\right)
    E^{\prime q}_{33}+\sigma^q_{12}E^{\prime q}_{23} \\[0.3cm]
E_{22}^{\prime q}\sigma^q_{21} & 0 & \sigma^q_{23}E^{\prime q}_{33}
\\[0.3cm]
E^{\prime
  q}_{33}\left(\sigma^q_{31}-\sigma^q_{32}\sigma^q_{21}\right)+E^{\prime
  q}_{32}\sigma^q_{21}\hspace{0.5cm} & E^{\prime q}_{33}\sigma^q_{32}
& 0
\end{array}} \right)\!.
\label{Etilde}
\end{eqnarray}
The fields $H^0_u$ and $H^0_d$ decompose into the physical components
$H^0$, $h^0$ and $A^0$ as
\begin{eqnarray}
H_u^0&=&\frac{1}{\sqrt{2}}\left(H^0\sin\alpha + h^0\cos\alpha +
iA^0\cos\beta\right)\,, \nonumber\\
H_d^0&=&\frac{1}{\sqrt{2}}\left(H^0\cos\alpha - h^0\sin\alpha +
iA^0\sin\beta\right)\,.  \label{Htilde}
\end{eqnarray}\medskip

Without the non-holomorphic corrections $E^{\prime q}_{ij}$ the
rotation matrices $U^{q\,L,R}$ would simultaneously diagonalize the
effective mass terms and the neutral Higgs couplings in \eq{L-Y-FCNC}.
However, in the presence of non-holomorphic corrections this is no
longer the case and apart from a flavor-changing non-holomorphic
correction also a term proportional to a flavor-conserving
non-holomorphic correction times a flavor-changing self-energy is
generated.\medskip

It is instructive to discuss this effect also in the full theory. The
two diagrams in Fig.~\ref{fig:eff_higgs_vertex} (both involving a
holomorphic $A$-terms) have opposite sign and cancel in the limit
$\mu,A^{\prime q}\to 0$.  However, in the presence of non-holomorphic
terms the cancellation is incomplete and a part proportional to
$1-\dfrac{1}{1+\epsilon_b\tan\beta}$ (for down-quarks) survives
(second term in \eq{Etilde}).  Even though this term is formally of
higher loop order, it is numerically relevant due to its chiral
enhancement.\medskip

The non-holomorphic parts of the fermion self-energy, as defined
in~\eq{HoloDeco}, can be extracted from Eqs.~(\ref{eq:gluinoSE}),
(\ref{neutralino_SE}) and (\ref{chargino-SE}). Note that the whole
chargino contribution is always non-holomorphic except for
$\cot\beta$\,-\,suppressed terms.  The same is true for the neutralino
contribution except for the pure bino part which decomposes in the
same way as the (dominant) gluino contribution (the latter given
already in~\cite{Crivellin:2010er}).\medskip

Using \eq{Etilde} and \eq{Htilde}, the effective Lagrangian in
\eq{L-Y-FCNC} leads to the following effective Higgs-fermion-fermion
Feynman rules\footnote{Note that some of the Higgs-quark-quark couplings are suppressed by a factor $\cos\beta$ or $\sin\alpha$ stemming from the Higgs mixing matrices. If one decides to keep these suppressed couplings, one should be aware of the fact that they receive proper vertex corrections in which the suppression factor does not occur and which are thus $\tan\beta$-enhanced with respect to the tree-level couplings. Such enhanced corrections to the coupling of $H^{\pm}$ to right-handed up-quarks are important for $b\to s \gamma$ \cite{Carena:2000uj,Degrassi:2000qf}. (see \ref{up-quark-non-holomorphic})} (note that the CKM matrix $V$ in the charged Higgs coupling
is the physical one):
\begin{eqnarray}
{\Gamma_{u_f u_i }^{H_k^0\,LR\,\rm{eff}} } &=& x_u^k\left( \frac{m_{u_i }}{v_u}
\delta_{fi} - \widetilde E_{fi}^{\prime u}\cot\beta \right) + x_d^{k\star}
\widetilde E_{fi}^{\prime u}\,, \nonumber\\[0.2cm]
{\Gamma_{d_f d_i }^{H_k^0\,LR\,\rm{eff} } } &=& x_d^k \left( \frac{m_{d_i
}}{v_d} \delta_{fi} - \widetilde E_{fi}^{\prime d}\tan\beta \right) +
x_u^{k\star}\widetilde E_{fi}^{\prime d} \,,\nonumber \\[0.2cm]
{\Gamma_{u_f d_i }^{H^\pm\,LR\,\rm{eff} } } &=& \sum\limits_{j = 1}^3
{\sin\beta\, V_{fj} \left( \frac{m_{d_i }}{v_d} \delta_{ji}-
  \widetilde{E}^{\prime d}_{ji}\tan\beta \right)\, }
\nonumber\\[0.2cm]
{\Gamma_{d_f u_i }^{H^ \pm\,LR\,\rm{eff} } } &=& \sum\limits_{j = 1}^3
{\cos\beta\, V_{jf}^{\star} \left( \frac{m_{u_i }}{v_u} \delta_{ji}-
  \widetilde{E}^{\prime u}_{ji}\tan\beta \right)\, }\,,
 \label{Higgs-vertices-decoupling}
\end{eqnarray}
where for $H^0_k=(H^0,h^0,A^0)$ the coefficients $x_q^{k}$ are given
by\footnote{In principle also the renormalization of the 
Higgs potential should be addressed. Our derivation of chirally enhanced flavor effects does not 
depend on the specific relations between Higgs self-couplings and their masses.
Since no chirally-enhanced effects occur in the Higgs sector, it is 
consistent to use the tree-level values for the Higgs parameters. 
However, one can as well use the NLO values for the Higgs masses and 
mixing angles which might be even better from the numerical point of view.}

\begin{equation}
x_d^k \, = \,\left(-\frac{1}{\sqrt{2}}\cos\alpha,\,\frac{1}{\sqrt{2}}\sin\alpha,
\,\frac{i}{\sqrt{2}}\sin\beta\right), \hspace{1cm}
x_u^k \, = \, \left(-\frac{1}{\sqrt{2}}\sin\alpha,\,-\frac{1}{\sqrt{2}}\cos\alpha,
\,\frac{i}{\sqrt{2}}\cos\beta\right)\,.
\end{equation}
It is important to keep in mind that the $\sigma^f_{ij}$ in
\eq{Etilde} must be calculated using the bare quantities ($Y^{f(0)}$
and $V^{(0)}$). \medskip

For the lepton case, the non-vanishing effective Higgs vertices read
\begin{eqnarray}
{\Gamma_{\ell_f \ell_i }^{H_k^0\,LR\,\rm{eff} } } &=& x_d^k \left(
\frac{m_{\ell_i }}{v_d} \delta_{fi} - \widetilde E_{fi}^{\prime
  \ell}\tan\beta \right) + x_u^{k\star}\widetilde E_{fi}^{\prime
  \ell}\,,\nonumber \\[0.2cm]
{\Gamma_{\nu_f \ell_i }^{H^\pm\,LR\,\rm{eff} } } &=& \sum\limits_{j = 1}^3
{\sin\beta\, V^{\textrm{PMNS}}_{fj} \left( \frac{m_{\ell_i }}{v_d} \delta_{ji}-
  \widetilde{E}^{\prime \ell}_{ji}\tan\beta \right)\, }\,.
\end{eqnarray}\medskip

\begin{nfigure}{t}
\includegraphics[width=0.8\textwidth]{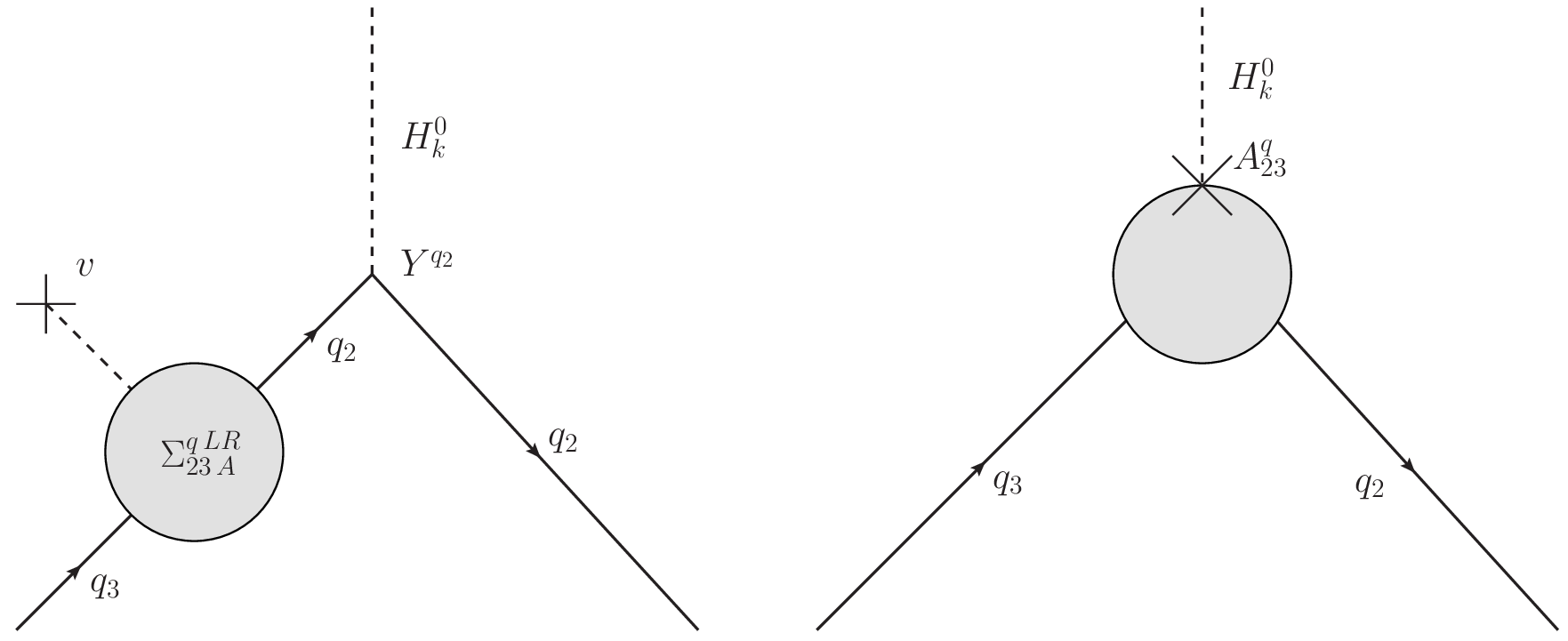}
\caption{Self-energy and genuine vertex correction involving
  $A^q_{23}$ contributing to the effective Higgs
  coupling.}\label{fig:eff_higgs_vertex}
\end{nfigure}

\section{Conclusions}
\label{sec:conclusions}

In the general MSSM, chirally-enhanced corrections are induced by
gluino-squark, chargino-sfermion and neutralino-sfermion loops and can
numerically compete with, or even dominate over, tree-level
contributions, due to their enhancement by either $\tan\beta$ or
$A^f_{ij}/(Y^f_{ij}\Msusy)$.  In this article we have identified all
potential sources of chirally-enhanced corrections and discussed their
effects on the finite renormalization of Yukawa couplings, fermion
wave-functions and the CKM matrix.  To leading order in $v/\Msusy$,
which numerically is a very good approximation for realistic choices
of MSSM parameters, we obtained analytic resummation formulae for
these quantities.\medskip

For the CKM resummation, it turned out to be useful to define a
generalized Wolfenstein parametrization, obtained by extending the
classical one to the case of complex $\lambda$ and $A$ parameters.
This parametrization is presented in~\ref{sec:wolf}.\medskip

For the resummation of the chirally-enhanced corrections in
supersymmetric fermion vertices, we have used the diagrammatic
approach developed in Refs.~\cite{Carena:1999py, Crivellin:2008mq,
  Hofer:2009xb, Crivellin:2009ar}.  This method allowed us to cast
chirally-enhanced corrections to gaugino(higgsino)-fermion-sfermion
couplings into effective vertices, as described in
Sec.~\ref{subsec:renseq} and~\ref{sec:effective-vertices}.\medskip

Moreover, we have given formulae for the effective
Higgs-fermion-fermion vertices, where we extended the results
of~\cite{Crivellin:2010er} by adding the chargino and neutralino
contributions.  Our effective Higgs-vertices can be used in the limit
$m_{H^0},m_{A^0},m_{H^{\pm}}\ll \Msusy$ as Feynman rules in an
effective theory with the SUSY particles being integrated
out. However, they remain still valid in the case
$m_{H^0},m_{A^0},m_{H^{\pm}}\sim \Msusy$ as long as the momenta
flowing through the Higgs vertices are much smaller than $\Msusy$.
Thus our effective Higgs-fermion-fermion Feynman rules can e.g. be
applied to calculate Higgs penguins contributing to
$B_{d,s}\to\mu^+\mu^-$, $B^+\to\tau^+\nu$ or the double Higgs penguin
contributing to $\Delta F=2$ processes.\medskip

If our effective matter fermion-sfermion-SUSY fermion and
Higgs-fermion-fermion Feynman rules are used for the calculation of an
Feynman amplitude at leading order in perturbation theory, all kinds
of chirally-enhanced effects are automatically included and resummed
to all orders in the result.

\begin{acknowledgments}
We are grateful to Ulrich Nierste for useful discussions and for a
careful proofreading of the manuscript.  A.C.~is supported by the
Swiss National Foundation. The Albert Einstein Center for Fundamental
Physics is supported by the ``Innovations- und Kooperationsprojekt
C-13 of the Schweizerische Universit\"atskonferenz SUK/CRUS''.  L.H.
is supported by the Federal Ministry of Education and Research (BMBF,
Germany) under contract No.~05H09WWE.  The work of J.R has been
supported in part by the Ministry of Science and Higher Education
(Poland) as research projects N N202 230337 (2009-12) and N N202
103838 (2010-12) and by the European Community's Seventh Framework
Programme under grant agreement PITN-GA-2009-237920 (2009-2013).
L.H. and J.R. like to thank the ITP Bern for the hospitality during
their visits there. We are grateful to Youichi Yamada, Christoph Borschensky and Jason Aebischer for bringing several typos in equations of the previous version of this article to our attention.
\end{acknowledgments}

\renewcommand{\thesection}{Appendix}
\renewcommand{\thesubsection}{\Alph{subsection}}
\numberwithin{equation}{subsection}

\section{}
\label{sec:app}

In this Appendix we collect definitions and conventions needed in the
article. Further, we define a generalized Wolfenstein parametrization
for the bare CKM matrix and give the non-holomorphic parts of the
up-quark self-energies with neutralions and charginos as virtual
particles.

\subsection{Generalized Wolfenstein Parametrization}
\label{sec:wolf}

While a general unitary $3\times 3$ matrix is described by 3 mixing angles and 6
complex phases, only one of those phases is physical in the case of
the CKM matrix $V$.  The other 5 phases are absorbed by proper
redefinition of the quark fields exploiting the $U(3)^ 3$-flavor
symmetry of the gauge interactions.  After application of this
procedure to the physical CKM matrix $V$, the quark field phases are
fixed.  As a consequence, possible additional phases in the bare CKM
matrix $V^{(0)}$, which originate from the diagonalization of the bare
Yukawa couplings $Y^{u(0)}$, $Y^{d(0)}$, cannot be absorbed anymore
and thus they are physical\footnote{In principle the additional phases
  could be absorbed into the wave functions of the bare quark fields
  $\psi^{(0)}$.  However, in this case they would modify the relation
  between the bare fields $\psi^{(0)}$ and the physical fields $\psi$
  and they would enter Feynman amplitudes in the form of
  complex CP-violating wave-function factors.  Since in this case CP
  violation in the quark sector would not be restricted to the CKM
  matrix anymore, we refrain from introducing this kind of
  wave-function rephasing.}.\medskip

The hierarchical structure of the measured CKM matrix $V$ can be made
explicit by using the Wolfenstein parametrization
\begin{equation}
\renewcommand{\arraystretch}{2.2} 
V \approx \left( {\begin{array}{*{20}c} 1 - \dfrac{\lambda^2}{2} &
    \lambda & {\lambda^3 A\left( {\rho - i\eta } \right)} \\
{ - \lambda } & {1 - \dfrac{{\lambda^2 }}{2}} & {\lambda^2 A} \\
{\lambda^3 A\left( {1 - \rho - i\eta } \right)} & { - \lambda^2 A} & 1
\\
\end{array}} \right)\label{Wolfenstein}
\end{equation}
with the small expansion parameter $\lambda=0.225$.  The three mixing
angles and the phase of $V$ are expressed via the four real parameters
$\lambda$, $A$, $\rho$, $\eta$.  Considering fine-tuning arguments, it
is reasonable to assume that the bare CKM matrix $V^{(0)}$ in the
general MSSM has a similar hierarchical structure as the physical CKM
matrix $V$.  Therefore it is desirable to have a parametrization
analogous to \eq{Wolfenstein} but allowing for possible additional
phases of $V^{(0)}$.\medskip

We will consider the following generalization of the Wolfenstein
parametrization:
\begin{equation}
\renewcommand{\arraystretch}{2.2} 
U\left( u_{12} ,u_{23}, u_{13}, u_{\rm{Im}} \right) \,=\, \left(
{\begin{array}{*{20}c} {1 - \dfrac{{\left| {u_{12} } \right|^2}}{2} +
      i u_{\rm{Im}} }\hspace{0.5cm} & { u_{12} } & { u_{13} } \\
{ - u_{12}^\star } & {1 - \dfrac{{\left| {u_{12}} \right|^2}}{2} - i
  u_{\rm{Im}} }\hspace{0.5cm & {u_{23} }} \\
{ - \left( {u_{13}^\star - u_{12}^\star u_{23}^\star } \right)} & { -
  u_{23}^\star } & 1 \\
\end{array}} \right).\label{generalized-Wolfenstein}
\end{equation}
The parameters $u_{12}\,,u_{23}\,,u_{13}\in\mathbb{C}$ and the
parameter $u_{\rm{Im}}\in\mathbb{R}$ should follow the hierarchical
structure of the usual Wolfenstein parametrization
(\ref{Wolfenstein}):
\begin{equation}
u_{12}={\cal O}(\lambda),\hspace{1.5cm} u_{23}={\cal
  O}(\lambda^2),\hspace{1.5cm} u_{13}={\cal
  O}(\lambda^3),\hspace{1.5cm} u_{\rm{Im}}={\cal O}(\lambda^2).
\end{equation}\medskip
Our parametrization is closed under hermitian conjugation and under
matrix multiplication.  We have (neglecting terms of ${\cal
  O}(\lambda^4)$ and higher)
\begin{itemize}
\item hermitian conjugation:
\begin{equation}
U^\dagger(u_{12},u_{23},u_{13},u_{\rm{Im}}) \,=\,
U(\widetilde{u}_{12},\widetilde{u}_{23},\widetilde{u}_{13},\widetilde{u}_{\rm{Im}})
\end{equation} 
where
\begin{eqnarray}
\widetilde{u}_{12}&=&-u_{12},\hspace{4.6cm}
\widetilde{u}_{23}\,=\,-u_{23},\nonumber\\
\widetilde{u}_{13}&=&-\left(u_{13}-u_{12}u_{23}\right),\hspace{2.5cm}
\widetilde{u}_{\rm{Im}}\,=\,-u_{\rm{Im}}\,.
\end{eqnarray}
\item matrix multiplication:
\begin{equation}
U(u_{12}^{\prime\prime},u_{23}^{\prime\prime},u_{13}^{\prime\prime},u_{\rm{Im}}^{\prime\prime})
\,=\, U(u_{12},u_{23},u_{13},u_{\rm{Im}}) \; \;
U(u_{12}^{\prime},u_{23}^{\prime},u_{13}^{\prime},u_{\rm{Im}}^{\prime})
\end{equation}
where
\begin{eqnarray}
u^{\prime\prime}_{12} &=& u_{12} + u^{\prime}_{12}, \hspace{3.1cm}
u^{\prime\prime}_{23} \,=\, u_{23} + u^{\prime}_{23},
\nonumber\\[0.2cm]
u^{\prime\prime}_{13} &=& u_{13} + u^{\prime}_{13} + u_{12}
u^{\prime}_{23},\hspace{1.5cm}
u^{\prime\prime}_{\rm{Im}} \,=\, u_{\rm{Im}} + u^{\prime}_{\rm{Im}} +
{\mathop{\rm Im}\nolimits} \left[ {u_{12} u_{12}^{\prime\star} }
  \right]\,.
\label{multiplication-rule}
\end{eqnarray}
\end{itemize}
Note in particular that the parameter $u_{\rm{Im}}$ had to be
introduced in order to make this parametrization closed under
multiplication.  \medskip

We will now demonstrate that our parametrization, which allows for 3
mixing angles and 4 complex phases, can be used to describe the bare
CKM matrix $V^{(0)}$ in the MSSM.  First we recognize that defining
$V=V\left(v_{12},v_{23},v_{13},v_{\rm{Im}}\right)$ with
\begin{equation}
  v_{12}=\lambda,\hspace{1.5cm} v_{23}=A\lambda^2, \hspace{1.5cm}
  v_{13}=\lambda^3 A\left( {\rho  - i\eta } \right), \hspace{1.5cm}
  v_{\rm{Im}}=0
\end{equation}
we recover the usual Wolfenstein parametrization (\ref{Wolfenstein}).
Furthermore, also the matrix $U^{f\,L}$ given in \eq{DeltaU} can be
described in the form $U^{f\,L} = U^{f\,L}(u^{f\,L}_{12},
u^{f\,L}_{23}, u^{f\,L}_{13},u^{f\,L}_{\rm{Im}})$:
\begin{equation}
u_{12}^{f\,L}=\sigma_{12}^f,\hspace{1.5cm}
u_{23}^{f\,L}=\sigma^f_{23}, \hspace{1.5cm}
u_{13}^{f\,L}=\sigma^f_{13}, \hspace{1.5cm} u_{\rm{Im}}^{f\,L}=0.
\end{equation}
Because the parametrization is closed under hermitian conjugation and
matrix multiplication, the relation between the physical CKM matrix
and the bare one in \eq{CKM-0-ren} implies that $V^{(0)}$ can also be
parametrized using \eq{generalized-Wolfenstein}.  In
Sec.~\ref{subsec:CKMreno} we took advantage of this parametrization of
$V^{(0)}$ in our study of the CKM renormalization.\bigskip

\subsection{Non-holomorphic part of the up-quark self-energy}
\label{up-quark-non-holomorphic}

The non-holomorpic part of the up-quark self-energy is not chirally enhanced 
and, since it therefore does not lead to large corrections to the Yukawa couplings, the CKM matrix or the fermion wave functions, it has been ommitted from 
eqs.~(\ref{eq:gluinoSE})-(\ref{chargino-SE}). Note, however, that its contribution to the effective Higgs-quark-quark couplings $\Gamma_{u_f u_i }^{H^0/A^0\,LR\,\rm{eff}}$ and $\Gamma_{d_f u_i }^{H^ \pm\,LR\,\rm{eff} }$ receives a relative $\tan\beta$ enhancement with respect to the $\cot\beta$-suppressed tree-level coupling (see \eq{Higgs-vertices-decoupling}). While the vertices $\Gamma_{u_f u_i }^{H^0/A^0\,LR\,\rm{eff}}$, $\Gamma_{d_f u_i }^{H^ \pm\,LR\,\rm{eff} }$ do not play a role for most phenomenological applications because of their $\cot\beta$-suppression, they are important for the Higgs contributions to $b\to s\gamma$. In the following we thus give formulae for the gluino-, neutralino- and chargino- contributions to the non-holomorphic part of the up-quark self-energy.\medskip 

The non-holomorphic part of the gluino contribution is easily obtained from \eq{eq:gluinoSE} by inserting $\Delta^{u\,LR}_{ij}\to -v_d A^{\prime u}_{ij}-v_d\,\mu\, Y^{u_i{(0)}}\, \delta_{ij}$. The non-holomorphic part of the neutralino contribution (including neutralino mixing) is given by
\bea
\Sigma _{u_i u_j}^{\tilde \chi _{}^0 \;LR} &=& \dfrac{-1}{16\pi^2}
\sum\limits_{m=1}^3 \left( - \dfrac{2}{9} g_1^2 M_1
\sum\limits_{i',j',n = 1}^3 \Lambda _{m\;ij'}^{u\;LL}
\Delta_{j'i'}^{u\;LR} \Lambda _{n\;i'j}^{u\;RR} B_0\left(
\left|M_1^2\right|, m_{\tilde q^L_m}^2, m_{\tilde u^R_n}^2 \right)
\right. \nonumber \\
&+& \left( \dfrac{g_1^2}{6}v_d M_1 \mu C_0 \left( \left|M_1^2\right|,
\left|\mu\right|, m_{\tilde q^L_m}^2 \right) - \dfrac{g_2^2 }{2}v_d
M_2 \mu C_0 \left( \left|M_2^2\right|, \left|\mu\right|, m_{\tilde
  q^L_m}^2 \right) \right)\Lambda_{m\;ij}^{u\;LL} Y^{u_j} \nonumber \\
&-& \left.  \dfrac{2}{3}g_1^2 v_d M_1 \mu Y^{u_i } \Lambda
_{m\;ij}^{u\;RR} C_0 \left( {\left| {M_1^2 } \right|,\left| \mu
  \right|,m_{\tilde u^R_m }^2 } \right) \right),
\eea
where again $\Delta^{u\,LR}_{ij}\to -v_d A^{\prime u}_{ij}-v_d\,\mu\, Y^{u_i{(0)}}\, \delta_{ij}$ must be substituted. Note that, as in the case of down-quarks, only the neutralino mixing induced by a coupling to the "wrong" Higgs gives a finite contribution while the holomorphic part which includes neutralino mixing would be divergent. The non-holomorphic part of the chargino contribution reads
\bea
\Sigma _{u_i u_j}^{\tilde \chi^\pm \;LR} &=& \dfrac{- 1}{16\pi^2}\mu
\!\!\! \sum\limits_{m,i',j'= 1}^3
\!\left(\sum\limits_{n,i^{\prime\prime},j^{\prime\prime} = 1}^3
\!\!V_{i'i}^{(0)} Y^{d_{i'}} \Lambda _{n\;i'i^{\prime\prime}}^{d\;RR}
\Delta_{i^{\prime\prime}j^{\prime\prime}}^{\tilde d\;RL} \Lambda
_{m\;j^{\prime\prime}j'}^{d\;LL} V_{j'j}^{(0)\star} Y^{u_j} C_0
\left( \left|\mu\right|^2, m_{\tilde q_m^L}^2, m_{\tilde d_n^R}^2
\right)\right. \nonumber \\
&-& \left. v_d g_2^2 M_2 \mu V_{i'i}^{(0)} \Lambda _{m\;i'j'}^{d\;LL}
V_{j'j}^{(0)\star} Y^{u_j } C_0 \left( {\left| {\mu ^2 }
  \right|,\left| {M_2^2 } \right|,m_{\tilde q_m^L }^2 } \right)
\right).
\eea
where one has to substitute $\Delta^{d\,LR}_{ij}\to -v_d A^{d}_{ij}$.\bigskip

\subsection{CKM renormalization in the case of CKM-dependent up-quark self-energies}
\label{upSE_CKM}

In the case of non-degenerate left-handed squark masses, the up-quark
self-energies depend on CKM elements due to the SU(2) relation between
the soft mass matrices of the left-handed squarks. The up-squark
mixing matrix $W^{u\,L}$ enters the gluino- and neutralino-
contributions to the up-quark self-energy through
$\Lambda^{u\,LL}_{m\,ij} = W^{u\,L}_{im} W^{u\,L\star}_{jm}$. The
SU(2) relation (\ref{SquarkMixSU2}) implies $\Lambda^{u\,LL}_{m\,fi} =
V^{(0)}_{fj}\Lambda^{q\,LL}_{m\,jk} V^{(0)\star}_{ik}$ and leads in
this way to a CKM-dependence of $\Sigma^{u\,LR}_{fi}$, which has been
made explicit in Eqs.~(\ref{eq:gluinoSE}), (\ref{neutralino_SE}). If
we assume that the off-diagonal elements $\Lambda^{q\,LL}_{m\,{fi}}$
are at most of the same order in the Wolfenstein parameter $\lambda$
as the corresponding elements $V_{fi}$ of the CKM matrix, we have to
leading order in $\lambda$:
\begin{eqnarray}
\Lambda_{m\,12}^{u\,LL} &=& \Lambda^{q\,LL}_{m\,12}\,+\,V_{12}^{\left(
  0 \right)}V^{(0)\star}_{22} \left( {\Lambda_{m\,22}^{q\,LL} -
  \Lambda_{m\,11}^{q\,LL} } \right)\,, \nonumber\\[0.2cm]
\Lambda_{m\,23}^{u\,LL} &=& \Lambda^{q\,LL}_{m\,23}\,+\,V_{23}^{\left(
  0 \right)}V^{(0)\star}_{33} \left( {\Lambda_{m\,33}^{q\,LL} -
  \Lambda_{m\,22}^{q\,LL} } \right)\,, \nonumber\\[0.2cm]
\Lambda_{m\,13}^{u\,LL} &=& \Lambda^{q\,LL}_{m\,13}\,+\,V_{13}^{\left(
  0 \right)}V^{(0)\star}_{33} \left( {\Lambda_{m\,33}^{q\,LL} -
  \Lambda_{m\,11}^{q\,LL} } \right) + V_{12}^{\left( 0 \right)}
V_{32}^{\left( 0 \right)\star} \left( {\Lambda_{m\,22}^{q\,LL} -
  \Lambda_{m\,11}^{q\,LL} } \right)\nonumber\\[0.2cm]
&&\,+\,V^{(0)}_{12}\Lambda^{q\,LL}_{m\,23}
V^{(0)\star}_{33}\,+\,V^{(0)}_{11}\Lambda^{q\,LL}_{12}V^{(0)\star}_{32}\,.
\end{eqnarray}
For the CKM renormalization it is important to distinguish between
contributions to $\Sigma^{u\,LR}_{fi}$ which depend on $V^{(0)}_{fi}$
and those which do not. To this end we decompose $\sigma^u_{fi}$ in
analogy to \eq{sigmahatd} for $\sigma^d_{fi}$ as
\begin{equation}
\sigma^u_{fi}\,=\,\widehat{\sigma}^u_{fi}\, +
\,V^{(0)}_{fi}V^{(0)\star}_{ii}\varepsilon^u_{fi}\,
\hspace{0.5cm}(f\not=i).
\end{equation}
For $i,f\not=3$ the quantity $\widehat{\sigma}^u_{fi}$ does not depend
on any off-diagonal CKM element. The parameters
$\widehat{\sigma}^u_{f3}$ and $\widehat{\sigma}^u_{3i}$ depend on the
CKM elements $V^{(0)}_{12}$ and $V^{(0)}_{23}$ (or equivalently on
$V^{(0)}_{21}$ and $V^{(0)}_{32}$), but they neither depend on
$V^{(0)}_{13}$ nor on $V^{(0)}_{31}$.  The parameters
$\varepsilon^u_{fi}$ are given by
\begin{eqnarray}
\varepsilon^u_{fi}&=&\frac{1}{\max\{m_{u_i},m_{u_f}\}}\sum\limits_{m,n=1}^3
\left(\Lambda^{q\,LL}_{m\,ii}-
\Lambda^{q\,LL}_{m\,ff}\right)\,\Delta^{u\,LR}_{33}\,\Lambda^{u\,RR}_{n\,33}
\nonumber\\
&&\times\,\left[\frac{2\alpha_s}{3\pi}m_{\tilde{g}}\,C_0\! \left(
  m_{\tilde{g}}^2,m_{\tilde q_m^L }^2,m_{\tilde u_n^R }^2 \right)
  \,+\,\frac{g_1^2}{72\pi^2}M_1\,C_0\! \left( \left| {M_1 }
  \right|^2,m_{\tilde q_m^L }^2,m_{\tilde u_n^R }^2 \right)\right]\,.
\end{eqnarray}
The term $\left(\Lambda^{q\,LL}_{m\,ii} -
\Lambda^{q\,LL}_{m\,ff}\right)$ causes a strong GIM suppression of
$\varepsilon^u_{fi}$ culminating in $\varepsilon^u_{fi}=0$ for
degenerate squark masses. Therefore it is a good approximation to
neglect higher-order effects related to $\varepsilon^u_{fi}$ in the
resummation formula for $V^{(0)}$, as it has been done in \eq{barCKM}
using the approximation (\ref{sigmahatu}). For completeness we derive
here an extended version of \eq{barCKM} resumming the effects of
$\varepsilon^u_{fi}$ to all orders.\medskip

We decompose the wave-function rotation matrix $U^{u\,L}$ as
\begin{equation}
U^{u\,L}\,=\,U^{u\,L}_{\CKM}\,\widehat{U}^{u\,L}\,\label{Udeco2}
\end{equation}
in analogy to \eq{Udeco} for the down sector.  The CKM-independent
part $\widehat{U}^{u\,L}$ is defined by replacing $\sigma_{ji}^u \to
\widehat{\sigma}^u_{ji}$ in \eq{DeltaU}, what amounts to the
generalized Wolfenstein parametrization
\begin{equation}
\widehat{U}^{u\,L}\,=\,\widehat{U}^{u\,L}\left(\widehat{\sigma}^u_{12},
\,\widehat{\sigma}^u_{23},\, \widehat{\sigma}^u_{13},\,0\right).
\end{equation} 
The CKM-dependent part $U^{u\,L}_{\CKM}$ is then given by
\begin{equation}
U^{u\,L}_{\CKM}\, = \,U^{u\,L}\,\widehat{U}^{u\,L\dag}\, =
\,U^{u\,L}_{\CKM}\left((u^{u\,L}_{\CKM})_{12},\,(u^{u\,L}_{\CKM})_{23},\,
(u^{u\,L}_{\CKM})_{13},\,(u^{u\,L}_{\CKM})_{\rm{Im}}\right)
\end{equation}
with
\begin{eqnarray}
  (u^{u\,L}_{\CKM})_{12}&=&V^{(0)}_{12}V^{(0)\star}_{22}\varepsilon^u_{12}\,,\hspace{2cm}
  (u^{u\,L}_{\CKM})_{23}\,=\,V_{23}^{(0)}V_{33}^{(0)\star}\varepsilon^u_{23}\,,\nonumber\\[0.2cm]
  (u^{u\,L}_{\CKM})_{13}&=&V^{(0)}_{13}V^{(0)\star}_{33}\varepsilon^u_{13}-V^{(0)}_{12}V^{(0)\star}_{22}\varepsilon^u_{12}\widehat{\sigma}^u_{23}\,,\nonumber\\[0.2cm]
  (u^{u\,L}_{\CKM})_{\rm{Im}}&=&-\rm{Im}\left[V^{(0)}_{12}V^{(0)\star}_{22}
    \varepsilon^u_{12}\widehat{\sigma}^{u\star}_{12}\right]\,.
\end{eqnarray}\medskip

Inserting the decomposition (\ref{Udeco}) and (\ref{Udeco2}) into
\eq{CKM-0-ren} we obtain
\begin{equation}
V^{(0)} \,=\, U^{u\,L}_{\CKM}\, \widetilde{V}\,U^{d\,L\dag}_{\CKM}\,.
   \label{CKM-reno3}   
\end{equation}
The matrix $\widetilde{V}$ is defined in \eq{VtildeDef} and its
elements are given in terms of generalized Wolfenstein parameters in
\eq{Vtilde2}. Solving \eq{CKM-reno3} for $V^{(0)}$, we finally get
\begin{eqnarray} 
 v_{12}^{(0)} &=& \dfrac{{\widetilde v_{12} }}{{1 -
     \varepsilon^{u}_{12} }}\,,\hspace{2cm} v_{23}^{(0)} =
 \dfrac{{\widetilde v_{23} }}{{1 - \varepsilon_{FC}^d -
     \varepsilon^{u}_{23} }}\,, \nonumber\\[0.2cm]
v_{13}^{\left( 0 \right)} &=& \frac{1}{{1 -
    \varepsilon _{FC}^d - \varepsilon _{13}^u  }}\left( {\widetilde v_{13} + \frac{{\widetilde v_{12}\varepsilon
      _{12}^u
      \left( {\widetilde v_{23} - \widehat \sigma _{23}^u } \right) }}{{1 - \varepsilon _{12}^u }} + 
      \frac{{\widetilde v_{12} \widetilde v_{23}
      \varepsilon _{FC}^d }}{{\left( {1 - \varepsilon _{12}^u }
      \right)\left( {1 -  \varepsilon _{FC}^d - \varepsilon _{23}^u }
      \right)}}} \right) \,, \nonumber\\[0.2cm]
v_{\rm{Im}}^{(0)}&=& \widetilde{v}_{\rm{Im}}\, +
\,\textrm{Im}\left[\dfrac{\widetilde{v}^{}_{12} \varepsilon^u_{12}
    \left(\widetilde{v}^{\star}_{12}-\widehat{\sigma}^{u\star}_{12}\right)}
  {1-\varepsilon^u_{12}}\right]\,.
 \label{CKMsigmaU}
\end{eqnarray}
For the application of \eq{CKMsigmaU} one has to keep in mind that
$\widehat{\sigma}^u_{13}$ depends on $V^{(0)}_{12}$ and
$V^{(0)}_{23}$. Therefore one has to proceed as follows: in a first
step $v_{12}^{(0)}$, $v_{23}^{(0)}$ and $v_{\rm{Im}}^{(0)}$ are
calculated from \eq{CKMsigmaU}. The results are used to determine
$\widehat{\sigma}^u_{13}$. With the help of $\widehat{\sigma}^u_{13}$
one can then calculate $\widetilde{v}_{13}$ from \eq{Vtilde2} and
finally $v_{13}^{(0)}$ from \eq{CKMsigmaU}.\bigskip

\subsection{Tree-level Feynman rules}
\label{sec:feyrul}

The tree level Feynman rules used throughout the paper are based on
those listed in Refs.~\cite{Rosiek:1989rs, Rosiek:1995kg}.  One should
however note few differences in conventions, which we summarize in
Table~\ref{tab:prd41}. Furthermore, the bare Yukawa couplings
calculated in Sec.~\ref{sec:Renormalization} are in general complex,
as shown explicitly in vertices displayed below. We use the convention
that $Y^{f(0)}$ is the coupling appearing in the $P_R$ component of
the (pseudo-)scalar Higgs-fermion-fermion vertex, whereas
$Y^{f(0)\star}$ appears in the $P_L$ component\footnote{Also Yukawa
  couplings in the LR blocks of the sfermion mass matrices, used to
  calculate sfermion mixing matrices, should be treated as complex. To
  find the correct positions of complex stars in the sfermion mass
  matrices, in our conventions one can use the mnemotechnic
  replacement rule $\mu Y\to \mu Y^{(0)},\mu^\star Y\to \mu^\star
  Y^{(0)\star}$.}.\medskip

\begin{table}[h]
\begin{center}
\begin{tabular}{|p{2mm}cp{2mm}|p{2mm}cp{2mm}|p{2mm}cp{2mm}|}
\hline
& Parameter &&& Current paper &&& Refs.~\cite{Rosiek:1989rs,
  Rosiek:1995kg} & \\
\hline
& Down-quark and lepton Yukawa couplings &&& $Y^\ell, Y^d$ &&& $-Y^\ell, -Y^d$ & \\
& Higgs vevs &&& $ \left\langle H_{u(d)} \right\rangle= v_{u(d)}$ &&&
$\left\langle H_{u(d)}\right\rangle = v_{u(d)}/\sqrt{2}$ & \\
& Lepton A-terms &&& $A^\ell_{ij}, A^{'\ell}_{ij}$ &&&
$-A^{\ell\star}_{ij}, A^{'\ell\star}_{ij}$ & \\
& Down-squark A-terms &&& $A^d_{ij}, A^{'d}_{ij}$ &&&
$-A^{d\star}_{ij}, A^{'d\star}_{ij}$ & \\
& Up-squark A-terms &&& $A^u_{ij},A^{'u}_{ij}$ &&&
$A^{u\star}_{ij},A^{'u\star}_{ij}$ & \\
& Squark mass terms &&& $\Delta^{f\,LL}_{ij}$, $\Delta^{f\,RR}_{ij}$, $\Delta^{f\,LR}_{ij}$ &&&
$\Delta^{f\,LL}_{ji}$, $\Delta^{f\,RR}_{ji}$, $\Delta^{f\,LR}_{ji}$ &\\
& Sfermion mass matrices &&& $\mathcal{M}_{f}^2$ &&&
$\left(\mathcal{M}_{f}^2\right)^\star=\left(\mathcal{M}_{f}^2\right)^T$ &
\\ \hline
\end{tabular}
\caption{Differences in conventions for the MSSM parameters in the
  current paper and in~\cite{Rosiek:1989rs, Rosiek:1995kg}.}
\label{tab:prd41}
\vspace*{-5mm}
\end{center}
\end{table}

Below we list the Feynman rules for the
gaugino(higgsino)-fermion-sfermion vertices.  The general definitions
of supersymmetric fermion and sfermion mixing matrices are given
in~\cite{Rosiek:1989rs, Rosiek:1995kg}. In \eq{eq:Wmat} we introduced
squark mixing matrices $W^{u,d}$ for the decoupling limit. These
matrices can be obtained from $Z_{U,D}$ in~\cite{Rosiek:1989rs,
  Rosiek:1995kg} substituting
\bea
Z_D = \left(
\begin{array}{cc}
W^{dL\star} & 0 \\
0 & W^{dR\star} \\
\end{array}
\right) + {\cal O}\left( \frac{v}{\Msusy}\right) 
\hskip 5mm
Z_U = \left(
\begin{array}{cc}
W^{uL} & 0 \\
0 & W^{uR} \\
\end{array}
\right) + {\cal O}\left( \frac{v}{\Msusy}\right) 
\label{eq:zdef}
\eea
Note the complex stars on the up-squark mixing matrices in
\eq{eq:zdef}, which have been added in order to stay compatible with
conventions of~\cite{Rosiek:1989rs, Rosiek:1995kg}.\bigskip\\

\begin{minipage}{0.25\textwidth}
\includegraphics[width=\textwidth]{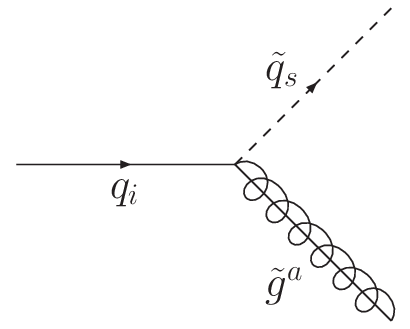} 
\end{minipage}
\hskip 2cm
\begin{minipage}{0.74\textwidth}
\lefteqn{
i\left[ \Gamma_{q_{i\beta} \tilde q_{s\alpha}}^{\tilde g L} P_L +
  \Gamma_{q_{i\beta} \tilde q_{s\alpha}}^{\tilde g R} P_R \right] \,
\rm{with}
}
\lefteqn{
\Gamma_{d_{i\beta} \tilde d_{s\alpha}}^{\tilde g L} = - g_s
\sqrt{2} T^a_{\alpha\beta} Z_D^{is}
}
\lefteqn{
\Gamma_{d_{i\beta} \tilde d_{s\alpha}}^{\tilde g R} = g_s \sqrt{2}
T^a_{\alpha\beta} Z_D^{i+3,s}
}
\lefteqn{
\Gamma_{u_{i\beta} \tilde u_{s\alpha}}^{\tilde g L} = - g_s
\sqrt{2} T^a_{\alpha\beta} Z_U^{is\star}
}
\lefteqn{
\Gamma_{u_{i\beta} \tilde u_{s\alpha}}^{\tilde g R} = g_s \sqrt{2}
T^a_{\alpha\beta} Z_U^{i+3,s\star}
}
\end{minipage}
\bigskip\\

\begin{minipage}{0.25\textwidth}
\includegraphics[width=\textwidth]{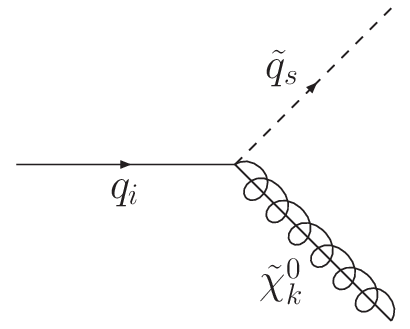} 
\end{minipage}
\hskip 2cm
\begin{minipage}{0.74\textwidth}
\lefteqn{
i\left[ {\Gamma_{q_i \tilde q_s }^{\tilde \chi_k^0 L} P_L + \Gamma
  _{q_i \tilde q_s }^{\tilde \chi_k^0 R} P_R } \right]\,{\rm{with}}
}
\lefteqn{ 
\Gamma_{d_i \tilde d_s }^{\tilde \chi_k^0 L} = \dfrac{1}{\sqrt{2}}
Z^{is}_{D} (g_2 Z^{2k}_{N} -\dfrac{1}{3}g_1Z^{1k}_{N}) - Y^{d_i(0)\star}
Z_{D}^{i+3,s} Z_{N}^{3k} 
}
\lefteqn{
\Gamma_{d_i \tilde d_s }^{\tilde \chi_k^0 R} = -
\dfrac{g_1\sqrt{2}}{3} Z_{D}^{i+3,s} Z_{N}^{1k\star} - Y^{d_i(0)}
Z_{D}^{is} Z_{N}^{3k\star}
}
\lefteqn{
\Gamma_{u_i \tilde u_s }^{\tilde \chi_k^0 L} = -\dfrac{1}{\sqrt{2}}
Z^{is\star}_{U} (g_2 Z^{2k}_{N} + \dfrac{1}{3} g_1 Z^{1k}_{N}) -
Y^{u_i(0)\star} Z_{U}^{i+3,s\star} Z_{N}^{4k}
}
\lefteqn{ 
\Gamma_{u_i \tilde u_s }^{\tilde \chi_k^0 R} =
\dfrac{2\sqrt{2}g_1}{3} Z_{U}^{i+3,s\star} Z_{N}^{1k\star} - Y^{u_i(0)}
Z_{U}^{is\star} Z_{N}^{4k\star}
}
\end{minipage}
\bigskip\\

\begin{minipage}{0.25\textwidth}
\includegraphics[width=\textwidth]{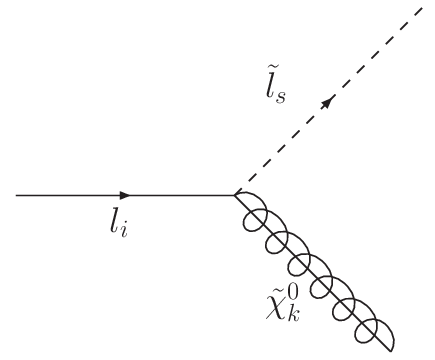} 
\end{minipage}
\hskip 2cm
\begin{minipage}{0.74\textwidth}
\lefteqn{ 
i\left[ {\Gamma_{\ell_i \tilde \ell_s }^{\tilde \chi_k^0 L} P_L +
    \Gamma_{\ell_i \tilde \ell_s }^{\tilde \chi_k^0 R} P_R }
  \right]\,{\rm{with}} 
}
\lefteqn{
\Gamma_{\ell_i \tilde \ell_s }^{\tilde \chi_k^0 L} =
\dfrac{1}{\sqrt{2}} Z_L^{is} (g_1 Z_N^{1k} + g_2 Z_N^{2k}) - Y^{\ell_i(0)\star}
Z_L^{i+3,s} Z_N^{3k}
}
\lefteqn{
 \Gamma_{\ell_i \tilde \ell_s }^{\tilde \chi_k^0 R} = -g_1\sqrt{2}
 Z_L^{i+3,s} Z_N^{1k\star} - Y^{\ell_i(0)} Z_{L}^{is} Z_{N}^{3k\star}
}
\end{minipage}
\bigskip\\

\begin{minipage}{0.25\textwidth}
\includegraphics[width=\textwidth]{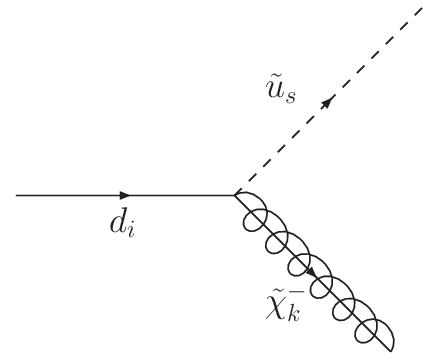} 
\end{minipage}
\hskip 2cm
\begin{minipage}{0.74\textwidth}
\lefteqn{
i\left[ {\Gamma_{d_i \tilde u_s }^{\tilde \chi_k^ \pm L} P_L +
    \Gamma_{d_i \tilde u_s }^{\tilde \chi_k^ \pm R} P_R }
  \right]\,{\rm{with}} 
}
\lefteqn{
\Gamma_{d_i \tilde u_s }^{\tilde \chi_k^ \pm L} = \sum\limits_{j =
  1}^3 ( - g_2 Z_U^{js\star} Z_+^{1k} + Y^{u_j(0)\star} Z_U^{(j+3)s\star}
Z_+^{2k}) V_{ji}^{(0)}
}
\lefteqn{
\Gamma_{d_i \tilde u_s }^{\tilde \chi_k^ \pm R} = Y^{d_i(0)}
\sum\limits_{j = 1}^3 Z_U^{js\star} Z_-^{2k\star} V_{ji}^{(0)} 
}
\end{minipage}
\bigskip\\

\begin{minipage}{0.25\textwidth}
\includegraphics[width=\textwidth]{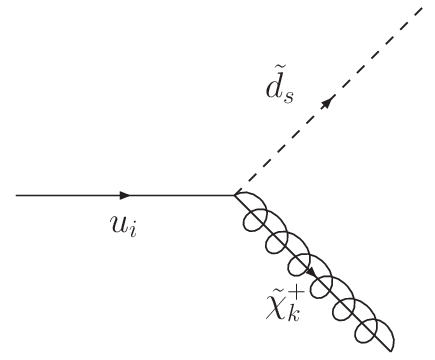} 
\end{minipage}
\hskip 2cm
\begin{minipage}{0.74\textwidth}
\lefteqn{
i\left[ {\Gamma_{u_i \tilde d_s }^{\tilde \chi_k^ \pm L} P_L +
    \Gamma_{u_i \tilde d_s }^{\tilde \chi_k^ \pm R} P_R }
  \right]\,{\rm{with}} 
}
\lefteqn{
\Gamma_{u_i \tilde d_s}^{\tilde \chi_k^ \pm L} = \sum\limits_{j =
  1}^3 ( - g_2 Z_D^{js} Z_-^{1k} + Y^{d_i(0)\star} Z_D^{j+3,s} Z_-^{2k})
V^{(0)\star}_{ij}
}
\lefteqn{
\Gamma_{u_i \tilde d_s }^{\tilde \chi_k^ \pm R} = \sum\limits_{j =
  1}^3 Y^{u_i(0)} Z_D^{js} Z_+^{2k\star} V^{(0)\star}_{ij}
}
\end{minipage}
\bigskip\\

\begin{minipage}{0.25\textwidth}
\includegraphics[width=\textwidth]{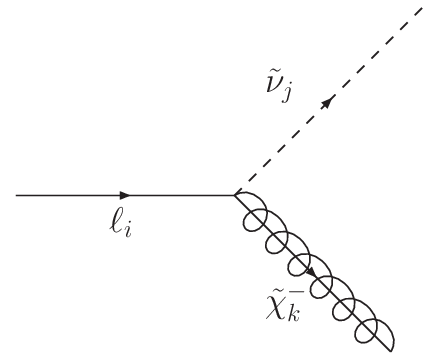} \end{minipage}
\hskip 2cm
\begin{minipage}{0.74\textwidth}
\lefteqn{
i\left[ {\Gamma_{\ell_i \tilde \nu_s }^{\tilde \chi_k^ \pm L} P_L +
    \Gamma_{\ell_i \tilde \nu_s }^{\tilde \chi_k^ \pm R} P_R }
  \right]\,{\rm{with}} 
}
\lefteqn{
\Gamma_{\ell_i \tilde \nu_s}^{\tilde \chi_k^ \pm L} = - g_2 Z_+^{1k}
Z_{\nu}^{is\star}
}
\lefteqn{
\Gamma_{\ell_i \tilde \nu_s }^{\tilde \chi_k^ \pm R} = Y^{\ell_i(0)}
Z_-^{2k\star} Z_{\nu}^{is\star}
}
\end{minipage}\bigskip
\bigskip\\

\begin{minipage}{0.25\textwidth}
\includegraphics[width=\textwidth]{./eps/lepton-sneutrino+chargino.eps} \end{minipage}
\hskip 2cm
\begin{minipage}{0.74\textwidth}
\lefteqn{
i\left[ {\Gamma_{\ell_i \tilde \nu_s }^{\tilde \chi_k^ \pm L} P_L +
    \Gamma_{\ell_i \tilde \nu_s }^{\tilde \chi_k^ \pm R} P_R }
  \right]\,{\rm{with}} 
}
\lefteqn{
\Gamma_{\ell_i \tilde \nu_s}^{\tilde \chi_k^ \pm L} = - g_2 Z_+^{1k}
Z_{\nu}^{is}
}
\lefteqn{
\Gamma_{\ell_i \tilde \nu_s }^{\tilde \chi_k^ \pm R} = Y^{\ell_i(0)}
Z_-^{2k} Z_{\nu}^{is}
}
\end{minipage}\bigskip

\subsection{Loop integrals}
\label{sec:loopint}

The momentum dependent loop functions in \eq{MSSM-self-energies} are
defined as 
\bea
B_0 \left( p^2 ;m_1^2 ,m_2^2 \right) &=& \dfrac{\left( {2\pi\mu}
  \right)^{4 - d} }{i\pi^2 }\int {d^d k} \dfrac{1}{(k^2 - m_1^2)\left(
  (k - p)^2 - m_2^2\right) }\,, \nn\\
p^\mu B_1 \left( p^2 ;m_1^2 ,m_2^2 \right) &=& \dfrac{\left( {2\pi \mu
  } \right)^{4 - d}}{i\pi^2}\int {d^d k} \dfrac{k^\mu}{(k^2 -
  m_1^2)\left( (k - p)^2 - m_2^2 \right)}\,.
\label{B_0-B_1}
\eea\medskip
Evaluating the function $B_0$ for vanishing external momentum, one gets
\begin{equation}
B_0 \left( m_1^2, m_2^2 \right) \,=\, B_0 \left( 0 ;m_1^2 ,m_2^2 \right)=1 + \dfrac{m_1^2 \ln
  \dfrac{Q^2}{m_1^2} - m_2^2 \ln \dfrac{Q^2}{m_2^2}}{m_1^2 - m_2^2}\,.
\end{equation}
Here a divergent constant $\frac{2}{4-d}-\gamma_{E}+\log 4\pi$ has
been dropped. It always cancels in the formulae of this article when
the sum over all internal particles is performed. The same is true for
the artificial scale $Q^2$.  The loop-functions $C_0$ and $D_0$ are
defined in analogy to $B_0$ but correspond to integrals with three and
four propagators, respectively. For vanishing external momenta they
are given by 
\bea 
C_0 \left( {m_1^2 ,m_2^2 ,m_3^2 } \right) &=& \dfrac{B_0 ( m_1^2
  ,m_2^2) - B_0 (m_1^2, m_3^2) }{m_2^2-m_3^2}\,, \nn \\[2mm]
&=& \dfrac{m_1^2 m_2^2 \ln \dfrac{m_1^2}{m_2^2} + m_2^2 m_3^2 \ln
  \dfrac{m_2^2}{m_3^2} + m_3^2 m_1^2 \ln \dfrac{m_3^2}{m_1^2}}{\left(
  m_1^2 - m_2^2 \right)\left( m_2^2 - m_3^2 \right) \left( m_3^2 -
  m_1^2 \right)}\,, \nn \\[2mm]
D_0 \left( {m_1^2 ,m_2^2 ,m_3^2,m_4^2 } \right) &=& \dfrac{C_0 ( m_1^2
  ,m_2^2,m_3^2) - C_0 ( m_1^2 ,m_2^2,m_4^2) }{m_3^2-m_4^2}\,.
\eea
\bigskip

\bibliography{chiral} 

\end{document}